# *A modal-Hamiltonian interpretation of quantum mechanics*♣

Olimpia Lombardi[a*] - Mario Castagnino[b]

[b] *CONICET-Universidad de Buenos Aires, Crisólogo Larralde 3440, 1430, Buenos Aires, Argentina.*

[a] *CONICET-IAFE, Universidad de Buenos Aires, Casilla de Correos 67, Sucursal 28, 1428, Buenos Aires, Argentina.*

**Abstract**

The aim of this paper is to introduce a new member of the family of the modal interpretations of quantum mechanics. In this *modal-Hamiltonian interpretation*, the Hamiltonian of the quantum system plays a decisive role in the property-ascription rule that selects the definite-valued observables whose possible values become actual. We show that this interpretation is effective for solving the measurement problem, both in its ideal and its non-ideal versions, and we argue for the physical relevance of the property-ascription rule by applying it to well-known physical situations. Moreover, we explain how this interpretation supplies a description of the elemental categories of the ontology referred to by the theory, where quantum systems turn out to be bundles of possible properties.

*Keywords*: Quantum mechanics; Modal interpretation; Hamiltonian; Quantum Measurement; Decoherence, Classical limit, Quantum Ontology.

**Contents**

# 1. Introduction

In one of its meanings, realism is the philosophical position according to which the aim of scientific theories is to describe reality. From this realist viewpoint, the purpose of an interpretation of a scientific theory is to say how reality would be if the theory were true.

During a long period after the first formulations of quantum mechanics, the orthodox interpretational framework was tied to an instrumentalist reading of the theory; from this perspective, quantum features were discussed only in terms of possible measurements results. But in the last decades, the traditional instrumentalist position has begun to loose its original strength, and several realist interpretations of quantum mechanics have been proposed. In this context, modal interpretations aim at assigning properties to physical systems on the basis of the quantum-mechanical formalism, without attributing a special role to measurement: quantum measurements are conceived as ordinary physical interactions, and measurements outcomes as properties of measurement apparatuses. Therefore, modal interpretations are *realist*, *non-collapse* interpretations, according to which the quantum state of a system describes the possible properties of the system rather than the properties that it actually possesses.

Modal interpretations can be viewed as a family, whose members differ to each other mainly in the specific rule designed to ascribe actually possessed properties to quantum systems. In this paper we want to propose a new member of this family: we have called it "*modal-Hamiltonian interpretation*" due to the central role played by the Hamiltonian both in the definition of quantum systems and subsystems and in the selection of the actual properties. We shall argue for the fruitfulness of this interpretation by stressing its physical relevance and by showing how it solves the difficulties into which some modal interpretations run when faced with non-ideal measurements. Moreover, we shall show that our actualization rule supplies a satisfactory answer to the problem of the classical limit of quantum mechanics. Finally, we shall discuss the ontological picture suggested by this interpretation, pointing out its philosophical foundations and its conceptual advantages.

## 2. Why modal?

The roots of the modal interpretations can be found in the works of Bas van Fraassen (1972, 1973, 1974), where the distinction between the *quantum state* and what he called the *value state* of the system is introduced: the quantum state tells us what may be the case, that is, which physical properties the system *may* possess; the value state represents what *actually* is the case. The relationship between the quantum state and the value state is *probabilistic*. Therefore, the quantum state is the basis for modal statements, that is, statements about what possibly or necessarily is the case.[1]

On the basis of this original idea, in the 1980s several authors presented realist interpretations which, in retrospect, can be regarded as elaborations or variations on van Fraassen's modal themes (for an overview and references, see Dieks & Vermaas, 1998). However, each one of them proposed its own rule of property-ascription.[2] For instance, in the so-called Kochen-Dieks modal interpretation (Kochen, 1985; Dieks, 1988), the biorthogonal (Schmidt) decomposition of the pure quantum state of the system picks out the definite-valued observables.[3] The Vermaas-Dieks version (Vermaas & Dieks, 1995), a generalization of the Kochen-Dieks interpretation to mixed states, is based on the spectral resolution of the reduced density operator: the range of the possible properties of a system and their corresponding probabilities are given by the non-zero diagonal elements of the spectral resolution of the system's reduced state, obtained by partial tracing. In turn, the so-called atomic modal interpretation (Bacciagaluppi & Dickson, 1999) is based on the assumption that there exists a special set of disjoint systems, which are the building blocks of all other systems, and that set fixes a preferred factorization of the Hilbert space; the properties of a system supervene on the properties ascribed to its "atomic" subsystems. More recently, and as a response to some difficulties in the account of non-ideal measurements, Gyula Bene and Dennis Dieks (2002) have developed a perspectival version of the modal interpretation, according to which properties are not monadic but have a relational character: following the original idea of Kochen (1985), systems have properties "as witnessed" by a larger system.

---

[1] Although van Fraassen does not endorse scientific realism (from his "constructive empiricism", the aim of science is only empirical adequacy), he admits that a meaningful account of reality is necessary for a scientific theory to be intelligible.

[2] Bub (1992, 1994) classifies modal interpretations on the basis of the preferred observables that acquire definite values in each one of them.

[3] There is, however, a relevant difference between Kochen's and Dieks' positions. For Kochen, the properties are not possessed by the system absolutely, but only when it is "witnessed" by another system. By contrast, for Dieks in those first works, the properties ascribed to the system do not have a relational character.

In this section we are not interested in the differences among the members of the modal family, but rather in the features that they share. In particular, all the modal interpretations agree on the following points (for a clear summary, see Dieks, 2007, Section 1):

- The interpretation is based on the standard formalism of quantum mechanics.
- The interpretation is realist, that is, it aims at describing how reality would be if quantum mechanics were true.
- Quantum mechanics is a fundamental theory, which must describe not only elementary particles but also macroscopic objects.
- Quantum mechanics describes single systems: the quantum state refers to a single system, not to an ensemble of systems.
- The quantum state of the system (pure state or mixture) describes the possible properties of the system and their corresponding probabilities, and not the actual properties (given by van Fraassen's value state). The relationship between the quantum state and the actual properties of the system is probabilistic.
- Systems possess actual properties at all times, whether or not a measurement is performed on them.
- A quantum measurement is an ordinary physical interaction. There is no collapse: the quantum state always evolves unitarily according to the Schrödinger equation.
- The Schrödinger equation gives the time evolution of probabilities, not of actual properties.

If these are the features that make an interpretation qualify as a member of the modal family, then our interpretation also belongs to this family. As we shall see, the main difference between our proposal and the previous modal versions consists in the central role played by the Hamiltonian of the quantum system in the selection of the actual properties.[4]

---

[4] In Section 8 we shall also take a definite position about the ontological picture arising from our interpretation. The question about the categories of the ontology referred to by quantum mechanics in the light of a particular interpretation has not been widely discussed in the literature about modal interpretations.

## 3. Structure and content of quantum mechanics

The first step of our proposal is the explicit presentation of what we shall consider the formal structure and the physical content of quantum mechanics. This step might seem superfluous for the specialist; however, if we want to build a new interpretation, we have to be clear about our theoretical commitments from the very beginning. This is particularly relevant due to the fact that our interpretation moves away from the traditional Hilbert space formulation of the theory, by adopting the algebraic formalism of quantum mechanics as a departing point. Although this might seem a minor formal detail, it will play a central role in the ontological picture arising from our interpretation (see Section 8). Moreover, since we shall argue for the physical relevance of our proposal (see Section 5), we have to show how the formal structure of the theory is endowed with a physical meaning.

Let us begin by stressing that we should distinguish between the mathematical language (self-adjoint operators, functionals, eigenstates and eigenvalues of an operator, etc.) and the physical language (observables, states, values of the observables, etc.): each term of the mathematical language refers to a physical item whose name belongs to the physical language. It is this *physical interpretation of the mathematical formalism* what turns it into the formalism of a *physical theory*. However, since this distinction would make the reading long and tedious, we shall follow the usual presentations, where both languages are mixed under the assumption that the reader knows the difference between mathematical and physical terms. Nevertheless, this difference will be explicit in the table included at the end of Section 4.

### 3.1 Formal structure

According to the algebraic formalism of quantum mechanics (see Haag, 1993), given a *-algebra $\mathcal{A}$ of operators, (i) the set of the self-adjoint elements of $\mathcal{A}$ is the space $\mathcal{O}$, whose elements represent observables, $O \in \mathcal{O}$, and (ii) states are represented by functionals on $\mathcal{O}$, that is, by elements of the dual space $\mathcal{O}'$, $\rho \in \mathcal{O}'$. In this theoretical framework, the observables are the basic elements of the theory, and states are secondary elements, defined in terms of the basic ones (we shall return on this point in Section 8).

We shall adopt a C*-algebra of operators to formulate the quantum postulates. As it is well known, a C*-algebra can be represented by a Hilbert space $\mathcal{H}$ (GNS theorem) and, in this particular case, $\mathcal{O} = \mathcal{O}'$; therefore, $\mathcal{O}$ and $\mathcal{O}'$ are represented by $\mathcal{H} \otimes \mathcal{H}$. The first quantum postulate gives the definition of a quantum system.

> **QP1:** A quantum system $S$ is defined by a pair $(\mathcal{O} \subseteq \mathcal{H} \otimes \mathcal{H}, H)$ where: (i) the observables $O$ of $S$ are represented by self-adjoint operators belonging to the space of observables $\mathcal{O}$, $O \in \mathcal{O} \subseteq \mathcal{H} \otimes \mathcal{H}$, and (ii) $H$ is a particular observable called Hamiltonian of $S$, $H \in \mathcal{O} \subseteq \mathcal{H} \otimes \mathcal{H}$.

This definition of quantum system (as well as the next quantum postulates), even if corresponding to the C*-algebra case, could be rephrased for different *-algebras under the necessary conditions for the representation of the algebra.[5]

> **QP2:** Given a quantum system $S: (\mathcal{O} \subseteq \mathcal{H} \otimes \mathcal{H}, H)$, and the observable $O = \sum_i o_i P_i \in \mathcal{O}$, where $o_i \neq o_j$ for $i \neq j$ and the $P_i \in \mathcal{O}$ are projector observables (eigenprojectors of $O$), each eigenvalue $o_i$ of $O$ is one of the possible values of $O$.

> **QP3:** Given a quantum system $S: (\mathcal{O} \subseteq \mathcal{H} \otimes \mathcal{H}, H)$, its initial state $\rho_0$ is represented by a self-adjoint functional belonging to $\mathcal{O}'$ (the dual space of $\mathcal{O}$) included in $\mathcal{H} \otimes \mathcal{H}$, $\rho_0 \in \mathcal{O}' \subseteq \mathcal{H} \otimes \mathcal{H}$, such that (i) $\text{Tr}(\rho_0)=1$ (normalization condition), and (ii) $\langle u | \rho_0 | u \rangle \geq 0$ for any $|u\rangle \in \mathcal{H}$ (non-negativeness condition).

> **QP4 (Dynamical Postulate):** Given a quantum system $S: (\mathcal{O} \subseteq \mathcal{H} \otimes \mathcal{H}, H)$ with initial state $\rho_0 \in \mathcal{O}'$, the state of $S$ at time $t$, $\rho(t) \in \mathcal{O}'$, is given by the Schrödinger equation (in its von Neumann version): $\rho(t) = e^{-iHt/\hbar} \rho_0 e^{iHt/\hbar}$.

The Born Rule introduces the probabilistic ingredient in the theory. In the instrumentalist readings of quantum mechanics, the rule is conceived as assigning probabilities to the results of measurements; this requires that "measurement" be conceived as a primitive concept of the theory. Since from our perspective measurement is an ordinary physical interaction that has to be accounted for by the theory with its interpretation, the Born Rule is endowed with a realist content: it assign probabilities to events in which observables acquire values. If we use the symbol $(A:a_k)$ to denote the event that an observable $A$ acquires the value $a_k$, the Born Rule can be expressed as follows.

---

[5] For instance, by means of a generalization of the GNS theorem (Iguri & Castagnino, 1999) it can be proved that a nuclear algebra can be represented by a rigged Hilbert space.

**QP5 (Born Rule):** Given a quantum system $S: (\mathcal{O} \subseteq \mathcal{H} \otimes \mathcal{H}, H)$ with initial state $\rho_0 \in \mathcal{O}'$, and considering an observable $A = \sum_i a_i P_i = \sum_i a_i \sum_{\mu_i} |i,\mu_i\rangle\langle i,\mu_i| \in \mathcal{O}$, where $a_i \neq a_j$ for $i \neq j$, the Born probability $P_B$ of the event $(A{:}a_k)$ at time $t$ can be computed as

$$P_B(A{:}a_k, \rho(t)) = Tr(\rho(t) P_k) = \sum_{\mu_k} \langle k,\mu_k | \rho(t) | k,\mu_k \rangle$$

where $P_k \in \mathcal{H} \otimes \mathcal{H}$ is the eigenprojector of $A$ that projects on the subspace spanned by all the $|k,\mu_k\rangle$ corresponding to $a_i = a_k$.

In particular, since the projectors $P \in \mathcal{H} \otimes \mathcal{H}$ ($P^2 = P$) represent observables with eigenvalues $p_1 = 1$ and $p_2 = 0$, the Born probability of the event $(P{:}1)$ results

$$P_B(P{:}1, \rho(t)) = Tr(\rho(t) P) \qquad (3\text{-}1)$$

If the Born probability $P_B$ is computed over all the projector observables $P \in \mathcal{H} \otimes \mathcal{H}$, the resulting probability function does not satisfy the definition of probability of Kolmogorov, since such a definition applies on a Boolean algebra whereas the set of the events corresponding to the projectors $P$ does not have a Boolean structure. For this reason, some authors define a generalized probability function (non-Kolmogorovian) over the orthoalgebra of quantum events (for instance, Hughes, 1989; Cohen, 1989). On the contrary, we shall preserve the Kolmogorovian character of probability by defining different probability functions on the Hilbert space $\mathcal{H}$. Let us define a *context* as a *complete set of orthogonal projectors* (*CSOP*) $\{\Pi_\alpha\}$, with $\alpha = 1$ to $M$, such that

$$\sum_{\alpha=1}^{M} \Pi_\alpha = I \quad \text{and} \quad \Pi_\alpha \Pi_\beta = \delta_{\alpha\beta} \Pi_\alpha \qquad (3\text{-}2)$$

where $I$ is the identity operator in $\mathcal{H} \otimes \mathcal{H}$.[6] All the projectors of the form $P = \sum_J \Pi_J$, where the $\Pi_J \in \text{CSOP} \{\Pi_\alpha\}$ and $J \in \mathcal{J} \subset \{1,\cdots,M\}$, define a Boolean algebra $B^\alpha = (\mathcal{P}^\alpha, \cap, \cup, \bar{\ }, 0, 1)$, where $\mathcal{P}^\alpha = \{x / x = \bigcup_J \{\Pi_J\}\}$. Then, through the Born Rule, a state $\rho \in \mathcal{O}'$ defines on $\mathcal{P}^\alpha$ a probability function $f_\rho^\alpha : \mathcal{P}^\alpha \to [0,1]$ that satisfies the axioms of Kolmogorov:

- $f_\rho^\alpha(\mathcal{U}) = 1$
- $f_\rho^\alpha(\varnothing) = 0$
- If $X, Y \in \mathcal{P}^\alpha$ and $X \cap Y = \varnothing$, then $f_\rho^\alpha(X \cup Y) = f_\rho^\alpha(X) + f_\rho^\alpha(Y)$

---

[6] It is clear that, if $\{|\alpha\rangle\}$ is a basis of the Hilbert space, the set $\{\Pi_\alpha = |\alpha\rangle\langle\alpha|\}$ is a CSOP.

As a consequence, given a CSOP $\{\Pi_\alpha\}$ and the projectors $P = \sum_J \Pi_J$:

- The probability $f_\rho^\alpha(P:1)$ ascribed to the event $(P:1)$ is computed with the Born Rule as

$$f_\rho^\alpha(P:1) = P_B(P:1, \rho) = Tr(\rho P) \tag{3-3}$$

- For any observable $A = \sum_i a_i P_i$, the probability $f_\rho^\alpha(A:a_k)$ ascribed to the event $(A:a_k)$ is identified with the probability $f_\rho^\alpha(P_k:1)$ ascribed to the event $(P_k:1)$:

$$f_\rho^\alpha(A:a_k) = f_\rho^\alpha(P_k:1) = Tr(\rho P_k) \tag{3-4}$$

Summing up, since Booleanity is retained in each context, we can say that the state $\rho$ defines a different probability function on each context, but each one of these probability functions satisfies the definition of Kolmogorov.

## 3.2 Physical content

In the previous subsection we have presented what is usually called "the formal structure of quantum mechanics." As Ballentine (1998) points out, although such a structure is a necessary basis for the formulation of the theory, it has by itself very little physical content. When concrete physical problems are to be solved, the relevant observables of the system, endowed with a clear physical meaning, have to be identified. Those observables are closely related to space-time transformations.

A continuous transformation $\mathcal{T}$ with parameter $s$ acts on observables and state vectors as

$$A \longrightarrow A' = U_s A U_s^{-1} \tag{3-5}$$

$$|\varphi\rangle \longrightarrow |\varphi'\rangle = U_s |\varphi\rangle \tag{3-6}$$

where $U_s$ is the family of unitary operators that describe the transformation $\mathcal{T}$. If $s$ is allowed to become infinitesimally small, the infinitesimal unitary operator $U_s$ can be expressed as

$$U_s = e^{iKs} \tag{3-7}$$

where $K$ is the *generator* of the transformation $\mathcal{T}$. In turn, when the state vector is represented as a function of space-time coordinates $x$, there is an inverse relation between transformations on function space and transformations on coordinates:

$$\varphi(x) = U_s \varphi(x') \tag{3-8}$$

As it is well known, each physical theory has a corresponding group of symmetry transformations, in the sense that the dynamical law of the theory is covariant under the transformations of the group.[7] In particular, the group corresponding to quantum mechanics is the Galilean group, defined by ten symmetry generators associated to ten parameters: one time-displacement, three space-displacements, three space-rotations and three boost-velocity components. Those symmetry generators represent the basic physical observables of the theory (see Ballentine, 1998): the energy $H$ (time-displacement), the momentum $P = (P_x, P_y, P_z)$ (space-displacement), the total angular momentum $J = (J_x, J_y, J_z)$ (space-rotation), and the position $Q = (Q_x, Q_y, Q_z)$ (boost-transformation, whose generator is $mQ$, where $m$ is the mass).[8] Moreover, the commutation relations for these observables can be obtained on the basis of the properties of the group: for instance, $[P_\alpha, P_\beta] = 0$, $[P_\alpha, H] = 0$, $[J_\alpha, J_\beta] \neq 0$, $[J_\alpha, H] = 0$, etc. The observables corresponding to the mass $M = mI$, where $I$ is the identity operator, the internal energy $W = H - P^2/2m$, and $S^2$, where $S = J - P \times Q$ is the spin angular momentum (or simply "spin"), are the Casimir operators of the group.[9]

As we have seen, there is an inverse relation between transformations on function space and transformations on coordinates (see eq.(3-8)). In the case of time-displacement, the transformation $x \to x'$ is $t_0 \to t_0 + \tau$, and $U_s$ is $U_\tau = e^{i(H/\hbar)\tau}$ since the Hamiltonian $H$ (strictly speaking, $H/\hbar$, see Note 8) is the generator of the transformation:

$$|\varphi(t_0)\rangle = e^{i(H/\hbar)\tau}|\varphi(t_0 + \tau)\rangle \qquad (3\text{-}9)$$

Making $t_0 = 0$ and $\tau = t$, we obtain

$$|\varphi(t)\rangle = e^{-i(H/\hbar)t}|\varphi(0)\rangle \qquad (3\text{-}10)$$

which has the form of the Schrödinger equation. However, it has to be noted that eq.(3-10) can be obtained only when the Hamiltonian $H$ is *independent of* $t$: if $H$ is a function of $t$, in general no simple closed form can be given for $U_\tau$ and, as a consequence, $H$ cannot be conceived as the generator of time-displacements (see Ballentine, 1989, p. 89). This means that the time-independence of the Hamiltonian is what endows the Schrödinger equation with a clear physical meaning (precisely, that of expressing time-displacements) and, at the same time, what makes the Schrödinger equation strictly applicable to *closed systems*. This result is what implicitly supports the orthodox formulation of quantum mechanics, where the quantum system is conceived as a closed, constant-energy system, unitarily evolving according to the Schrödinger equation; the Hamiltonian (the energy of the system) only changes with time as the result of the interaction with other systems, described by means of an interaction Hamiltonian. Our interpretation will be based on this orthodox version of the theory.[10]

---

[7] We are using the following notion of covariance. Let $Q_1, \ldots, Q_n$ be quantities that are functions of space-time coordinates, and $Q'_1, \ldots, Q'_n$ the result of applying a certain group of transformations. A physical law $L(Q_1, \ldots, Q_n) = 0$ is covariant under that group if it preserves its form under the transformations of the group: $L(Q'_1, \ldots, Q'_n) = 0$.

[8] Strictly speaking, the generators are proportional to these observables with a factor $1/\hbar$. For instance, the time-displacement generator is $K_\tau = H/\hbar$.

[9] The Galilean group is a Lie group, and an operator commuting with all the elements of a Lie group is said to be the Casimir operator of the group (see Tung, 1985). Since covariance under boost-transformatons holds, a quantum system with constant velocity can be described in the frame of reference of the center of mass, where $P = 0$; in this case, $W = H$, and the Hamiltonian becomes a Casimir operator of the Galilean group.

Although in a –closed– quantum system the Hamiltonian $H$ is time-independent and, then, invariant under time-displacements, it may have the remaining space-time symmetries or not. To say that the Hamiltonian is invariant under a certain symmetry transformation with generator $K$ and parameter $s$ means that (see eq.(3-5))

$$e^{iKs} H e^{-iKs} = H, \quad \text{then} \quad [H,K] = 0 \qquad (3\text{-}11)$$

This implies that, when $H$ is invariant under a certain continuous transformation, the generator of that transformation is a *constant of motion* of the system: each symmetry of the Hamiltonian defines a conserved quantity. For instance, the invariance of $H$ under space-displacements in any direction implies that the momentum $P$ is a constant of motion; the invariance of $H$ under space-rotations in any direction implies that the total angular momentum $J$ is a constant of motion. If, on the contrary, $H$ is invariant under space-displacements only in one direction, say $x$, only the component $P_x$ of $P$ is a constant of motion.

In turn, we know that the invariance under space-time transformations follows from the properties of space and time: invariance under time-displacements expresses the homogeneity of time, invariance under space-displacements expresses the homogeneity of space, invariance under space-rotations expresses the isotropy of space. We also know that the Schrödinger equation has to be covariant under the full Galilean group. The covariance of the Schrödinger equation under time-displacements is guaranteed by the time-independence of the Hamiltonian, and this expresses the homogeneity of time. However, in quantum mechanics, fields are not quantized; they are treated as classical fields that act on the quantum system by breaking the homogeneity and/or the isotropy of space. But, in spite of this, the dynamical law of the theory, $|\varphi(t)\rangle = e^{-iHt/\hbar}|\varphi(0)\rangle$, must remain covariant under space-displacements and space-rotations.

---

[10] An example of a non-orthodox approach is the position of Nancy Cartwright (1983), who proposes to consider the non-unitary generalized master equation for open quantum systems as the dynamical postulate of quantum mechanics instead of the Schrödinger equation.

Therefore, the breaking of the homogeneity and/or the isotropy of space resulting from the action of fields on the quantum system has to be "contained" in the form of the Hamiltonian: the non-homogeneity of space implies the non-invariance of $H$ under space-displacements and, then, the fact that $P$ (or one of its components) is not a constant of motion; the non-isotropy of space implies the non-invariance of $H$ under space-rotations and, then, the fact that $J$ (or one of its components) is not a constant of motion. Summing up, the privileged status of the Hamiltonian follows from its role in the Schrödinger equation: $H$ has to embody the space asymmetries in each particular situation in order to preserve the covariance of the dynamical law of the theory.

As we have seen, the study of the space-time symmetry transformations serves a dual purpose: the identification of the fundamental physical magnitudes of the theory, and the explanation of the central role played by the Hamiltonian in the selection of the constants of motion of the system. In other words, space-time symmetries endow the formal skeleton of quantum mechanics with the physical flesh and blood that allow the theory to be applied to concrete physical situations.

## 4. The modal-Hamiltonian interpretation

As we have pointed out, from a realist perspective, to interpret a theory amounts to saying how reality would be if the theory were true. Although, in general, physicists agree in their use of the physical language, it is not a self-evident matter what the relation between physical language and reality is: physical theories do not provide their own interpretations. Therefore, if we want to give an interpretation for quantum mechanics, we have to formulate *interpretational postulates* which define the ontological reference of each one of the theoretical elements introduced by the quantum postulates. In other words, we have to specify which kind of items in the ontology (objects, properties, facts, etc.) is represented by each physical term in the postulates QP (systems, observables, states, etc.). In this way we shall be able to say what *ontological categories* populate the quantum mechanical reality. For this reason, although we have been not very careful in the mathematical-physical language distinction, we shall pay a special attention to the ontological language: the task of fixing *the ontological reference of the physical language* is unavoidable if we want to understand the picture of reality supplied by our interpretation.

In this section, we shall introduce our modal-Hamiltonian interpretation without discussing its advantages over other proposals. The arguments that support it will become clear in the following sections, where we shall argue for its physical relevance and we shall apply it to solve some traditional interpretational challenges.

### 4.1 Properties, systems and subsystems

The first step consists in identifying the basic elements of the quantum ontology by means of interpretational postulates.

> **IP1:** Given a quantum system $S:(\mathcal{O} \subseteq \mathcal{H} \otimes \mathcal{H}, H)$, the observables $O \in \mathcal{O}$ ontologically represent *type-properties* $[O]$, and their corresponding eigenvalues $o_i$ ontologically represent *case-properties* $[O:o_i]$ of the type-property $[O]$. In particular, the projectors $P \in \mathcal{O}$ are observables that ontologically represent type-properties $[P]$ with case-properties $[P:1]$ and $[P:0]$.

Of course, any quantum system can be decomposed into parts in many ways; however, not any decomposition leads to parts which are, in turn, quantum systems. This will only be the case when the components' behaviors are dynamically independent to each other, and this dynamical independence follows from the non-interaction among them (see Harshman & Wickramasekara, 2007a, b). On this basis, we shall adopt the following interpretational postulates.

> **IP2 (System Decomposition):** A quantum system $S:(\mathcal{O} \subseteq \mathcal{H} \otimes \mathcal{H}, H)$ with initial state $\rho_0 \in \mathcal{O}'$ is composite when two quantum systems $S^1:(\mathcal{O}^1 \subseteq \mathcal{H}^1 \otimes \mathcal{H}^1, H^1)$ and $S^2:(\mathcal{O}^2 \subseteq \mathcal{H}^2 \otimes \mathcal{H}^2, H^2)$ can be defined, such that
> (i) $\mathcal{O} = \mathcal{O}^1 \otimes \mathcal{O}^2 \subseteq \mathcal{H} \otimes \mathcal{H}$, where $\mathcal{H} = \mathcal{H}^1 \otimes \mathcal{H}^2$, and
> (ii) $H = H^1 \otimes I^2 + I^1 \otimes H^2 \in \mathcal{O}$, where $I^1$ and $I^2$ are the identity operators in $\mathcal{H}^1 \otimes \mathcal{H}^1$ and $\mathcal{H}^2 \otimes \mathcal{H}^2$ respectively.
> In this case, the initial states $\rho_0^1 \in \mathcal{O}^{1\prime}$ and $\rho_0^2 \in \mathcal{O}^{2\prime}$ of $S^1$ and $S^2$, respectively, are obtained as $\rho_0^1 = Tr_{(2)}\rho_0$ and $\rho_0^2 = Tr_{(1)}\rho_0$, where $Tr_{(i)}\rho_0$ is a partial trace of $\rho_0$, that is, the operation that traces over the Hilbert space of $S^i$.

In this case we shall say that $S^1$ and $S^2$ are *subsystems* of the *composite system* $S = S^1 \cup S^2$. When a quantum system is not composite, we shall call it *elemental system*. Therefore, the System Decomposition postulate supplies a precise criterion to distinguish between elemental and composite systems, and such a criterion is *based on the system's Hamiltonian*.

It is worth stressing that this definition of composite system does not imply that the initial state $\rho_0$ of $S$ is the tensor product $\rho_0^1 \otimes \rho_0^2$: this factored or uncorrelated state is a very special kind of state, used in practice to describe independently prepared systems (see IP3 below). On the contrary, in general the initial state is a correlated or entangled state $\rho_0 \in \mathcal{O}'$; nevertheless, since there is no interaction between $S^1$ and $S^2$, $\left[H^1 \otimes I^2, I^1 \otimes H^2\right] = 0$ and, then,

$$exp\left[-iHt/\hbar\right] = exp\left[-iH^1t/\hbar\right]exp\left[-iH^2t/\hbar\right] \quad (4\text{-}1)$$

Therefore,

$$\rho^1(t) = Tr_{(2)}\rho(t) = Tr_{(2)}\left[e^{-iHt/\hbar}\rho_0 e^{iHt/\hbar}\right] = e^{iH^1t/\hbar}\left[Tr_{(2)}\rho_0\right]e^{-iH^1t/\hbar} = e^{iH^1t/\hbar}\rho_0^1 e^{-iH^1t/\hbar}$$
(4-2)

$$\rho^2(t) = Tr_{(1)}\rho(t) = Tr_{(1)}\left[e^{-iHt/\hbar}\rho_0 e^{iHt/\hbar}\right] = e^{iH^2t/\hbar}\left[Tr_{(1)}\rho_0\right]e^{-iH^2t/\hbar} = e^{iH^2t/\hbar}\rho_0^2 e^{-iH^2t/\hbar}$$
(4-3)

This means that, in spite of the correlations, the subsystems $S^1$ and $S^2$ are *dynamically independent*: each one of them will evolve under the action of its own Hamiltonian.

Harshman and Wickramasekara (2007a, b) call "tensor product structure" (TPS) any partition of the whole system $S$, represented in the Hilbert space $\mathcal{H} = \mathcal{H}^A \otimes \mathcal{H}^B$, into parts represented in $\mathcal{H}^A$ and $\mathcal{H}^B$. They point out that quantum systems admit a variety of TPSs, each one of which leads to a different entanglement between those parts: separability and entanglement are TPS-dependent. However, they also note that there is a particular TPS that is dynamically invariant and, as a consequence, the resulting entanglement is also dynamically invariant: that particular TPS corresponds to our decomposition of the *composite* system into *subsystems*. Therefore, when $S^1$ and $S^2$ are subsystems of the composite system $S$ and, then, dynamically independent, we have a robust notion of entanglement: although $\rho_0$ evolves in time, its entanglement is dynamically invariant.

> **IP3 (System Composition):** Given two quantum systems $S^1: (\mathcal{O}^1 \subseteq \mathcal{H}^1 \otimes \mathcal{H}^1, H^1)$ and $S^2: (\mathcal{O}^2 \subseteq \mathcal{H}^2 \otimes \mathcal{H}^2, H^2)$, with initial states $\rho_0^1 \in \mathcal{O}^1{}'$ and $\rho_0^2 \in \mathcal{O}^2{}'$ respectively, a quantum system $S: (\mathcal{O} \subseteq \mathcal{H} \otimes \mathcal{H}, H)$ with initial state $\rho_0 \in \mathcal{O}'$ can always be defined, such that
> (i) $\mathcal{O} = \mathcal{O}^1 \otimes \mathcal{O}^2 \subseteq \mathcal{H} \otimes \mathcal{H}$, where $\mathcal{H} = \mathcal{H}^1 \otimes \mathcal{H}^2$,
> (ii) $H = H^1 \otimes I^2 + I^1 \otimes H^2 + H_{int} \in \mathcal{O}$, where $H_{int}$ is usually called *interaction Hamiltonian*, and
> (iii) $\rho_0 = \rho_0^1 \otimes \rho_0^2 \in \mathcal{O}'$.

Although we have called IP3 "System Composition" postulate, two cases have to be distinguished:

- If $H_{int} = 0$ ($S^1$ and $S^2$ do not interact), $S^1$ and $S^2$ are subsystems of the composite system $S = S^1 \cup S^2$: they satisfy the System Decomposition postulate IP2. In this case, the initial states $\rho_0^1 = Tr_{(2)}\rho_0 = Tr_{(2)}\left(\rho_0^1 \otimes \rho_0^2\right)$ and $\rho_0^2 = Tr_{(1)}\rho_0 = Tr_{(1)}\left(\rho_0^1 \otimes \rho_0^2\right)$ evolve unitarily according to the Schrödinger equation, as required by the Dynamical Postulate QP4 (see eqs.(4-2) and (4-3)).

- If $H_{int} \neq 0$ ($S^1$ and $S^2$ interact), since the time $t_0$ when the interaction begins, $S^1$ and $S^2$ are not subsystems of the system $S$: they do not satisfy the System Decomposition postulate IP2. In fact, the initial states $\rho_0^1 = Tr_{(2)}\rho_0$ and $\rho_0^2 = Tr_{(1)}\rho_0$ do not evolve unitarily. Therefore, strictly speaking, since that time $t_0$, $S^1$ and $S^2$ are not quantum systems to the extent that they do not satisfy the Dynamical Postulate QP4.

   **IP4:** Given a composite quantum system $S = S^1 \cup S^2 : (\mathcal{O} \subseteq \mathcal{H} \otimes \mathcal{H}, H)$, where $S^1: (\mathcal{O}^1 \subseteq \mathcal{H}^1 \otimes \mathcal{H}^1, H^1)$ and $S^2: (\mathcal{O}^2 \subseteq \mathcal{H}^2 \otimes \mathcal{H}^2, H^2)$, and given the observables $A^1 \in \mathcal{O}^1$ of $S^1$, $A^2 \in \mathcal{O}^2$ of $S^2$, and the observables $A = A^1 \otimes I^2 \in \mathcal{O}$ and $A^f = f\left(A^1 \otimes I^2, I^1 \otimes A^2\right) \in \mathcal{O}$ of $S$, where $f$ is an analytical function, then,

   (i) the observables $A$ and $A^1$ ontologically represent *the same type-property* $[A] = [A^1]$ with the same case-properties $[A:a_i^1] = [A^1:a_i^1]$, where the $a_i^1$ are the eigenvalues of both $A$ and $A^1$.

   (ii) the observable $A^f$ ontologically represents a type-property $[A^f]$ with case-properties $[A^f : f(a_i^1, a_j^2)]$, where the $a_i^1$, $a_j^2$ are the eigenvalues of $A^1$ and $A^2$ respectively; $[A^f]$ is *equivalent* to the combination between $[A^1 \otimes I^2]$ and $[I^1 \otimes A^2]$, represented by the function $f$.

The interpretational postulate IP4 expresses the usual quantum assumption according to which the observable $A^1$ of a subsystem $S^1$ and the observable $A = A^1 \otimes I^2$ of the composite system $S = S^1 \cup S^2$ represent the same property. On the other hand, this postulate establishes the necessary connections between the properties of the composite system and the properties of its subsystems. The assumption of these connections is not a specific feature of quantum mechanics, but is also usual in classical mechanics where we consider, for instance, the energy of a two-particles composite system as a particular combination (expressed by the sum) of the energies of the component subsystems.

## 4.2 Possible facts and propensities

Up to this point we have identified *properties* in the ontology. But properties are not facts: properties and facts belong to different ontological categories. In our case, *possible facts* will be defined as the possible occurrence of certain case-properties, and probabilities will be applied to them.

**IP5:** Given a quantum system $S: (\mathcal{O} \subseteq \mathcal{H} \otimes \mathcal{H}, H)$,

(i) if $P \in \mathcal{O}$ is a projector observable, the *possible fact* $\langle\langle F[P] \rangle\rangle$ is the possible occurrence of the case-property $[P:1]$ corresponding to the type-property $[P]$.

(ii) if $A = \sum_i a_i P_i \in \mathcal{O}$, where $a_i \neq a_j$ for $i \neq j$ and the $P_i \in \mathcal{O}$ are the eigenprojectors of $A$, the *possible fact* $\langle\langle F[A:a_k] \rangle\rangle$ is the possible occurrence of the case-property $[A:a_k]$ corresponding to the type-property $[A]$.

**IP6:** Given an elemental quantum system $S: (\mathcal{O} \subseteq \mathcal{H} \otimes \mathcal{H}, H)$, a CSOP $\{\Pi_\alpha\}$, with $\alpha = 1$ to $M$, and the projectors $P = \sum_J \Pi_J \in \mathcal{O}$, where $\Pi_J \in \{\Pi_\alpha\}$ and $J \in \mathcal{J} \subset \{1, \cdots, M\}$, then

(i) the *possible fact* $\langle\langle F[P] \rangle\rangle$ is *equivalent* to the disjunction of the possible facts $\langle\langle F[\Pi_J] \rangle\rangle$:

$$\langle\langle F[P] \rangle\rangle \equiv \vee_J \langle\langle F[\Pi_J] \rangle\rangle$$

(ii) if $A = \sum_i a_i P_i \in \mathcal{O}$, where $a_i \neq a_j$ for $i \neq j$ and the $P_i \in \mathcal{O}$ are the eigenprojectors of $A$, the *possible fact* $\langle\langle F[A:a_k] \rangle\rangle$ is *equivalent* to the possible fact $\langle\langle F[P_k] \rangle\rangle$:

$$\langle\langle F[A:a_k] \rangle\rangle \equiv \langle\langle F[P_k] \rangle\rangle$$

It is clear that, given a CSOP $\{\Pi_\alpha\}$, the projector observables $P = \sum_J \Pi_J \in \mathcal{O}$ lead to a *set of possible facts* $\mathcal{F}^\alpha = \{x / x = \langle\langle F[P] \rangle\rangle\}$ that is isomorphic to the set $\mathcal{P}^\alpha = \{x / x = \bigcup_J \{\Pi_J\}\}$; then, $\mathcal{F}^\alpha$ also defines a Boolean algebra. As a consequence, in each elemental quantum system the probability function $f_\rho^\alpha$, defined by the state $\rho$ over the set $\mathcal{P}^\alpha$, can be ontologically interpreted as a Kolmogorovian measure of an ontological propensity, defined by the state $\rho$ over the set $\mathcal{F}^\alpha$.

**IP7:** Given an elemental quantum system $S: (\mathcal{O} \subseteq \mathcal{H} \otimes \mathcal{H}, H)$ in a state $\rho \in \mathcal{O}'$, and the set of possible facts $\mathcal{F}^\alpha$ defined by the CSOP $\{\Pi_\alpha\}$, with $\alpha = 1$ to $M$, the probability function $f_\rho^\alpha : \mathcal{P}^\alpha \to [0,1]$ ontologically represents a measure of *the ontological propensity to actualization* of the corresponding possible facts, $p_\rho^\alpha : \mathcal{F}^\alpha \to [0,1]$, a measure that satisfies the axioms of Kolmogorov expressed in terms of the Boolean connectives $\neg$, $\wedge$ and $\vee$.

As a consequence, given a CSOP $\{\Pi_\alpha\}$ and the projectors $P = \sum_J \Pi_J$:

- For a possible fact $\langle\langle F[P] \rangle\rangle$ (see eq.(3-3)):

$$p_\rho^\alpha(\langle\langle F[P] \rangle\rangle) = f_\rho^\alpha(P{:}1) = Tr(\rho P) \tag{4-4}$$

- For a possible fact $\langle\langle F[A{:}a_k] \rangle\rangle \equiv \langle\langle F[P_k] \rangle\rangle$ (see eq.(3-4)):

$$p_\rho^\alpha(\langle\langle F[A{:}a_k] \rangle\rangle) = f_\rho^\alpha(A{:}a_k) = Tr(\rho P_k) \tag{4-5}$$

**IP8:** Given an elemental quantum system $S{:}(\mathcal{O} \subseteq \mathcal{H} \otimes \mathcal{H}, H)$, its state $\rho \in \mathcal{O}'$ *codifies the ontological propensities to actualization* of the possible facts corresponding to the properties of $S$, and the time evolution of $\rho$ given by the Schrödinger equation ontologically represents *the time evolution of those ontological propensities*.

### 4.3 Actual facts

Up to this point we have talked of possible facts with their corresponding propensities to actualization. Now we have to introduce the ontological category of *actual facts*, resulting from the actualization of possible facts.

**IP9:** The *actual fact* $F[P]$ (or $F[A{:}a_k]$) results from the actualization of the possible fact $\langle\langle F[P] \rangle\rangle$ (or $\langle\langle F[A{:}a_k] \rangle\rangle$) and, therefore, it is the *actual occurrence* of the case-property $[P{:}1]$ (or $[A{:}a_k]$) corresponding to the type-property $[P]$ (or $[A]$).

With this terminology, we can read:
- $F[\cdot]$: "the possible fact $\langle\langle F[\cdot] \rangle\rangle$ is actual"
- $\neg F[\cdot]$: "the possible fact $\langle\langle F[\cdot] \rangle\rangle$ is not actual"

The theorem of Kochen and Specker (1967) teaches us that, at a single time $t$, the actualization of the possible facts belonging to all the sets $\mathcal{F}^\alpha$, defined by all the CSOP's of the system, is not permitted by the theory. Therefore, actualization must be restricted in some way; we shall introduce such a restriction by means of the following interpretational postulate.

**IP10:** Given an elemental quantum system $S:(\mathcal{O} \subseteq \mathcal{H} \otimes \mathcal{H}, H)$ such that $H \neq 0$,[11] there is a *preferred context*, that is, a *preferred CSOP* $\{\Pi_\alpha^p\}$ that determines a preferred set of possible facts $\mathcal{F}^p = \{x / x = \langle\langle F[P] \rangle\rangle\}$ (with $P = \sum_J \Pi_J^p \in \mathcal{O}$, $\Pi_J^p \in \{\Pi_\alpha^p\}$ and $J \in \mathcal{J} \subset \{1, \cdots, M\}$) where actualization occurs, in the sense that:

(i) one and only one of the possible facts $\langle\langle F[\Pi_\alpha^p] \rangle\rangle \in \mathcal{F}^p$, say $\langle\langle F[\Pi_\Omega^p] \rangle\rangle$, is also an actual fact $F[\Pi_\Omega^p]$.

(ii) if $F[\Pi_\Omega^p]$, then all the possible facts $\langle\langle F[P] \rangle\rangle \equiv V_J \langle\langle F[\Pi_J^p] \rangle\rangle \in \mathcal{F}^p$ (see IP6 (i)), such that $\Pi_\Omega^p$ is one of the $\Pi_J^p$, are also actual facts $F[P]$.

(iii) if $F[P_k]$, then the possible fact $\langle\langle F[A:a_k] \rangle\rangle \equiv \langle\langle F[P_k] \rangle\rangle \in \mathcal{F}^p$ (see IP6 (ii)) is also an actual fact $F[A:a_k]$.

Since quantum mechanics is a probabilistic theory, in general it does not determine which possible facts become actual. The actualization or non actualization of a possible fact belonging to the preferred set $\mathcal{F}^p$ is determined only when the measure of the propensity to actualization of such a fact is 1 or 0, respectively.

**IP11:** Given an elemental quantum system $S:(\mathcal{O} \subseteq \mathcal{H} \otimes \mathcal{H}, H)$ and the preferred set $\mathcal{F}^p$ of possible facts, if at time $t$ the state $\rho$ ascribes a propensity to actualization with measure 1 (measure 0) to the possible fact $\langle\langle F[\bullet] \rangle\rangle \in \mathcal{F}^p$, then such a possible fact is (is not) actual:

$$\text{If } p_\rho^p(\langle\langle F[\bullet] \rangle\rangle) = 1, \text{ then } F[\bullet]$$
$$\text{If } p_\rho^p(\langle\langle F[\bullet] \rangle\rangle) = 0, \text{ then } \neg F[\bullet]$$

### 4.4 Actualization rule

In the interpretational postulates IP10 and IP11, the preferred context has been presupposed. Now we have to supply a rule for identifying it: this task is usually the most delicate step of any modal interpretation. In fact, since quantum mechanics does not account for actualization due to its probabilistic nature, the rule cannot be inferred from the formalism, but has to be introduced as an interpretational assumption. Therefore, the adequacy of the Actualization Rule has to be assessed in the light of its physical relevance and its ability to solve the interpretation problems of the theory. But before introducing the rule, we want to discuss the conceptual motivations lying behind its adoption.

---

[11] The Hamiltonian equal to zero or equal to a multiple of the identity are physically equivalent situations, since they only differ in a time-displacement: $H = kI$ only modifies the phase of a state vector. So, for simplicity, from now on we shall only mention the case $H = 0$ on the basis of that equivalence.

During the last decades, the research on the mathematical properties of the formal structure of quantum mechanics has shown a great advance: many results, unknown by the founding fathers of the theory, have been obtained, and this work has greatly improved the understanding of the deep obstacles that any interpretation must face. However, this interest in the features of the formalism has led to forget the physical content of the theory: in the last times, usually the arguments rely on mathematical results and discussions center around the formal models of the quantum measurement. But quantum mechanics is a *physical* theory that has been applied to many well-known systems and by means of which an impressive amount of experimental evidence has been accounted for. Therefore, a "good" interpretation of quantum mechanics should not only face the traditional interpretational challenges of the theory, but also show its agreement with the orthodox practice of physics. In this sense, our proposal moves away from the present trend in the subject by placing an element with a clear physical meaning, the Hamiltonian of the system, at the heart of the interpretation: the content of the Actualization Rule will be based on the structure of the Hamiltonian.

Although in the previous section we have only considered invariances under space-time transformations, the Hamiltonian may have other symmetries. Precisely, the so-called "Schrödinger group" is the group of transformations that leave the Hamiltonian invariant (see Tinkham, 1964). Since each symmetry of the Hamiltonian leads to an energy degeneracy (Meijer & Bauer, 2004, pp. 110-114), much valuable information on the energy spectrum of the system can be obtained by applying the machinery of the group theory to the study of the symmetries of the Hamiltonian. In fact, if $H$ is invariant under a symmetry transformation with generator $K$, then $[H,K]=0$ (see eq.(3-11)). Therefore, if $H|n\rangle = \omega_n |n\rangle$,

$$K H |n\rangle = K \omega_n |n\rangle \quad \Rightarrow \quad H K |n\rangle = \omega_n K |n\rangle \tag{4-6}$$

This means that any vector $K|n\rangle$ obtained by applying the operator $K$ to the eigenvector $|n\rangle$ is also an eigenvector of $H$ with the same eigenvalue $\omega_n$. If $H$ is expressed as $H = \sum_n \omega_n P_n$, where $P_n$ is the eigenprojector corresponding to the eigenvalue $\omega_n$, we can write explicitly the index $k_n$ corresponding to the degeneracy of $\omega_n$ in such a way that

$$H|n,k_n\rangle = \omega_n |n,k_n\rangle \quad \Rightarrow \quad H = \sum_n \omega_n \sum_{k_n} |n,k_n\rangle\langle n,k_n| \tag{4-7}$$

$$K|n,k_n\rangle = \kappa_{k_n} |n,k_n\rangle \quad \Rightarrow \quad K = \sum_n \sum_{k_n} \kappa_{k_n} |n,k_n\rangle\langle n,k_n| \tag{4-8}$$

The degeneracies with origin in symmetries are called "normal" (Tinkham, 1964) or "systematic" (Cohen-Tannoudji, Diu & Lalöe, 1977). On the contrary, degeneracies that have no obvious origin in symmetries are called "accidental". However, deeper study usually shows either that the accidental degeneracy is not exact, or else that a hidden symmetry in the Hamiltonian can be found which explains the degeneracy.[12] For this reason it is assumed that, once all the symmetries of the Hamiltonian have been considered, a basis for the Hilbert space of the system is obtained and the "good quantum numbers" are well defined. This strategy is what underlies the group approach to quantum mechanics, where the physical features of the quantum system are studied by analyzing the symmetry properties of its Hamiltonian (Weyl, 1950; Hamermesh, 1962; Tinkham, 1964; Tung, 1985).[13]

Now we have all the conceptual elements necessary to present our Actualization Rule. The basic idea can be expressed by the classical Latin maxim "*Ubi lex non distinguit, nec nos distinguere debemus"*: *w*here the law does not distinguish, neither ought we to distinguish. The Hamiltonian of the system, with its symmetries, is what rules actualization; then, none observable whose eigenvalues would distinguish among eigenvectors corresponding to a single degenerate eigenvalue of the Hamiltonian has to acquire definite values, since the actualization of the possible facts corresponding to that observable would introduce in the system an asymmetry not contained in the Hamiltonian. Once this basic idea has been clearly understood, the Actualization Rule can be formulated in a very simple way.

> **IP12 (Actualization Rule):** Given an elemental quantum system $S: (\mathcal{O} \subseteq \mathcal{H} \otimes \mathcal{H}, H)$,
> (i) if $H = 0$, there is no actualization, that is, none possible fact becomes actual.
> (ii) if $H = \sum_n \omega_n P_n \neq 0$, the preferred CSOP $\{\Pi_\alpha^p\}$ is $\{P_n\}$, where each $P_n$ is an eigenprojector of $H$, which projects onto the subspace spanned by the eigenvectors corresponding to $\omega_n$.

---

[12] A classical example is the degeneracy, in the hydrogen atom, of states of different angular momentum $l$ but the same principal quantum number $n$, for instance, of $2s$ and $2p$ functions. In this case, Fock (1935) showed that the degeneracy arises from a four-dimensional rotational symmetry of the Hamiltonian in momentum space.

[13] More general group approaches supply generic mathematical strategies for identifying the symmetries of the Hamiltonian. Precisely, the quantum system is defined by a group $\mathcal{G}$ generated by the algebra of observables $\mathcal{A}$. If the subgroup chains $\mathcal{S}^\alpha \equiv \mathcal{G}_1^\alpha \supset \mathcal{G}_2^\alpha \supset \cdots \supset \mathcal{G}_n^\alpha$ of the group $\mathcal{G}$ are defined, the symmetries of the Hamiltonian are manifested by the fact that $H$ can be expressed in terms of the Casimir operators $C_{ki}^\alpha$ of any subgroup chain, where $C_{ki}^\alpha$ is the $i-th$ Casimir operator of the subgroup $\mathcal{G}_k^\alpha$ of the chain $\mathcal{S}^\alpha$ (see Barut & Raczka, 1987; Zhang, Feng, Yuan & Wang, 1989).

In other words, since the degeneracies of the Hamiltonian define a sort of "coarse-grained basis" of the Hilbert space, actualization must not introduce a finer discrimination that would break the symmetries due to the Hamiltonian.

In the modal interpretations discussed in the literature, the problem of identifying the actualization context is usually posed in terms of deciding which observables are definite-valued. Let us see how the rule works in those terms, for different cases:

(a) The Hamiltonian $H$ does not have symmetries: it is non-degenerate. In this case,

$$H|n\rangle = \omega_n |n\rangle \qquad \text{with } \omega_n \neq \omega_{n'} \tag{4-9}$$

where $\{|n\rangle\}$ is a basis of the Hilbert space $\mathcal{H}$. Since in this case the preferred CSOP is $\{|n\rangle\langle n|\}$, we can call $\{|n\rangle\}$ "preferred basis". Therefore, $n$ is the only good quantum number: the definite valued observables of the system are $H$ and all the observables commuting with $H$.

(b) The Hamiltonian $H$ has certain symmetries that lead to energy degeneracy. In this case, $H$ can be written as

$$H|n,i_n\rangle = \omega_n |n,i_n\rangle \quad \Rightarrow \quad H = \sum_n \omega_n \sum_{i_n} |n,i_n\rangle\langle n,i_n| = \sum_n \omega_n P_n \tag{4-10}$$

where $\omega_n \neq \omega_{n'}$ and the index $i_n$ expresses the degeneracy of the energy eigenvalue $\omega_n$. Let us consider an observable of the form

$$A = \sum_{n,i_n} a_n |n,i_n\rangle\langle n,i_n| = \sum_n a_n \sum_{i_n} |n,i_n\rangle\langle n,i_n| = \sum_n a_n P_n \tag{4-11}$$

where $a_n \neq a_{n'}$. It is clear that $[H, A] = 0$. Moreover, $A$ has the same degeneracy as $H$ since they have the same eigenprojectors: the subspace spanned by the degenerate eigenvectors corresponding to $a_n$ is the same as that spanned by the degenerate eigenvectors corresponding to $\omega_n$. In other words, $A$ has the same symmetries as $H$. Therefore, all the observables $A$ commuting with $H$ and having the form of eq.(4-11) are definite-valued. On the contrary, for instance, observables of the form

$$B = \sum_{n,i_n} b_{n,i_n} |n,i_n\rangle\langle n,i_n| \tag{4-12}$$

in spite of commuting with $H$, do not acquire definite values, since any of the actual facts $F\left[B : b_{n,i_n}\right]$ would discriminate among the eigenvectors corresponding to a single degenerate eigenvalue $\omega_n$ of $H$: they would distinguish where the Hamiltonian does not distinguish.

(c) An interesting particular case arises when all the eigenvalues $\omega_n$ have the same $i$-fold degeneracy: the index $i$, that expresses the energy degeneracy, is not a function of $n$. Then, in this case eq.(4-10) becomes

$$H|n,i\rangle = \omega_n |n,i\rangle \qquad (4\text{-}13)$$

As a consequence, the Hamiltonian can be decomposed as

$$H = \sum_n \omega_n \sum_i |n,i\rangle\langle n,i| = \sum_n \omega_n |n\rangle\langle n| \otimes \sum_i |i\rangle\langle i| = H^{ND} \otimes I^D \qquad (4\text{-}14)$$

This decomposition expresses the partition of the original system $S$ into two non-interacting subsystems $S^{ND}$ and $S^D$:

* The system $S^{ND}$ is represented in the Hilbert space $\mathcal{H}^{ND}$, with basis $\{|n\rangle\}$, and its Hamiltonian $H^{ND}$ is non-degenerate.
* The system $S^D$ is represented in the Hilbert space $\mathcal{H}^D$, with basis $\{|i\rangle\}$, and its Hamiltonian is $H^D = 0$.

Therefore, the original system $S$ is a composite system $S = S^{ND} \cup S^D$ such that

$$\mathcal{H} = \mathcal{H}^{ND} \otimes \mathcal{H}^D \ , \quad H = H^{ND} \otimes I^D + I^{ND} \otimes H^D + H_{int} = H^{ND} \otimes I^D \qquad (4\text{-}15)$$

where $H^D = H_{int} = 0$. As a consequence, the Actualization Rule has to be applied to each elemental subsystem:

* In $S^{ND}$ the preferred basis is $\{|n\rangle\}$: the definite-valued observables are $H^{ND}$ and all the observables belonging to $\mathcal{H}^{ND} \otimes \mathcal{H}^{ND}$ and commuting with $H^{ND}$.
* In $S^D$ there is no actualization because $H^D = 0$: the observables of $S^D$ do not acquire definite-values.

Summing up, in general we can say that, according to the Actualization Rule, the definite-valued observables of the system are the Hamiltonian $H$, and the observables commuting with $H$ and having, at least, the same symmetries as $H$. It is clear that our interpretation follows the lines of standard modal interpretations in the sense of dropping the so-called "eigenstate-eigenvalue link", according to which an observable possesses a definite value if and only if the system's state is an eigenvector of that observable.

Up to this point, by means of our Actualization Rule we have identified the observables that receive definite values in elemental quantum systems. It is easy to see that, when the elemental system is a subsystem of a composite system, the definite-valued observables of the composite system can also be identified. In fact, according to IP4, given a composite quantum system $S = S^1 \cup S^2$, if the observable $A^1 \in \mathcal{O}^1$ of $S^1$ acquires the definite value $a_\Omega^1$, the same happens for the observable $A = A^1 \otimes I^2 \in \mathcal{O}$ of $S$, since $A^1$ and $A$ represent the same type-property $[A^1] = [A]$ with the same case-properties $[A^1:a_i^1] = [A:a_i^1]$, where the $a_i^1$ are the eigenvalues of both $A^1$ and $A$. Moreover, also according to IP4, if the observable $A^1 \in \mathcal{O}^1$ of $S^1$ acquires the definite value $a_\Omega^1$ and the observable $A^2 \in \mathcal{O}^2$ of $S^2$ acquires the definite value $a_\Phi^2$, any observable $A^f = f(A^1 \otimes I^2, I^1 \otimes A^2) \in \mathcal{O}$ of $S$, representing the type-property $[A^f]$ with case-properties $[A^f : f(a_i^1, a_j^2)]$, is also definite-valued and acquires the definite value $f(a_\Omega^1, a_\Phi^2)$.

In some modal interpretations (e.g. Kochen-Dieks, Vermaas-Dieks), in general the rule of property-ascription does not assign the same value to $A^1$ and $A = A^1 \otimes I^2$; this fact contradicts the usual physical assumption –based on the observational indistinguishability of $A^1$ and $A$– according to which both observables represent the same property. This trouble has led to many discussions (see Dieks & Vermaas, 1998, pp. 109-115), and to different strategies directed to avoid it (see, for instance, the perspectival version of Bene & Dieks, 2002). In our interpretation the trouble does not arise since that usual assumption is imposed as an interpretational postulate (IP4). But it must be noted that this postulate can be formulated because we have a precise criterion to distinguish between elemental systems and composite systems. Without such a criterion, the definition of subsystems and composite systems would be relative to the arbitrary partition (to the particular TPS, in terms of Harshman and Wickramasekara, 2007a, b) chosen in each case; therefore, the Actualization Rule (or the rule for property-ascription in other modal interpretations) should apply to any quantum system, both subsystems and composite systems, and the above difficulty would arise. In our case, on the contrary, the precise criterion to distinguish between elemental systems and composite systems is given by IP2. Once the elemental quantum systems have been univocally identified, there is no problem in applying the Actualization Rule only to them and picking up the definite-valued observables of the composite system by means of the interpretational postulate IP4.

It is worth stressing that, in contrast with other interpretations, the preferred context where actualization occurs is not a function of time, since it depends on the Hamiltonian: the definite-valued observables always commute with the Hamiltonian and, therefore, they are constants of motion of the system. In other words, the observables that receive definite values are the same during all the "life" of the quantum system as such –precisely, as a closed system–: there is no need of accounting for the dynamics of the actual properties (we shall return on this point in Subsection 8.3).

Our Actualization Rule is not based on purely mathematical results (e.g. Schmidt theorem, spectral theorem); on the contrary, it bestows a central role on the Hamiltonian of the system. In fact, the rule is inspired by physical motivations that will become clear in the next section, when it will be applied to well-known physical models.

| Mathematics | Physics | Ontology |
|---|---|---|
| Self-adjoint operators $O \in \mathcal{O} \subseteq \mathcal{H} \otimes \mathcal{H}$ | Observables $O$ | Type-properties $[O]$ |
| Eigenvalues $o_i$ of $O$ $O\|i,\mu_i\rangle = o_i \|i,\mu_i\rangle$ | Values $o_i$ of the observable $O$ | Case-properties $[O{:}o_i]$ |
| | $O$ may acquire the value $o_i$ | Possible fact $\langle\langle F[O{:}o_i] \rangle\rangle$ |
| Probability function corresponding to the CSOP $\{\Pi_\alpha\}$ $f_\rho^\alpha : \mathcal{P}^\alpha \to [0,1]$ | Physical probability that the observable $O$ acquire the value $o_k$: $f_\rho^\alpha(O{:}o_k)$ | Ontological propensity to actualization of a possible fact: $p_\rho^\alpha(\langle\langle F[O{:}o_k] \rangle\rangle)$ |
| Functionals $\rho \in \mathcal{O}' \subseteq \mathcal{H} \otimes \mathcal{H}$ | States $\rho$ | Codification of propensities |
| Actualization is not accounted for by the theory | | Actual fact $F[O{:}o_k]$ |

## 5. The physical relevance of the interpretation

As we have said, a "good" interpretation of quantum mechanics should not only face the traditional interpretational challenges of the theory, but also show its agreement with the orthodox practice of physics. In this section we shall argue for the physical relevance of our modal-Hamiltonian interpretation by applying it to very well-known models and experimental results.

### 5.1 Free particle

The Hamiltonian of the free particle reads

$$H = \frac{P^2}{2m} = \frac{P_x^2 + P_y^2 + P_z^2}{2m} \tag{5-1}$$

where $P$ is the momentum observable, with components $P_x, P_y, P_z$, and $m$ is the mass of the particle. The particle is said to be "free" because there are not fields acting on it: then, space is homogeneous and, as a consequence, $H$ is invariant under space-displacements in any direction (an analogous argument could be given in terms of the isotropy of space). The components $P_x, P_y, P_z$ are the generators of the symmetry and, at the same time, constants of motion of the system. Therefore, the Hamiltonian is degenerate.

According to our Actualization Rule, $H$ acquires a definite value, and also $P^2$ since it is proportional to $H$ and, then, has the same space-displacement symmetry ($P^2$ is the Casimir operator of the group generated by $P_x, P_y, P_z$). Nevertheless, $P_x, P_y, P_z$ are not definite-valued because, being the generators of the symmetry, the actualization of any of the possible facts corresponding to their eigenvalues would break the symmetry of the free particle, in the sense of introducing an asymmetry non contained in the Hamiltonian.

Of course, the three components $P$ can be used for the theoretical description of the free particle; in fact, usually any two of them are added to $H$ to constitute a complete set of commuting observables (CSCO), $\{H, P_x, P_y\}$, $\{H, P_y, P_z\}$ or $\{H, P_x, P_z\}$, that defines a basis of the Hilbert space (given the functional dependence among the four magnitudes, the CSCO $\{P_x, P_y, P_z\}$ can be equivalently used). But this fact does not mean that those observables have to be considered definite-valued. On the contrary, the application of our interpretation to this system agrees with the non empirical accessibility to the values of $P_x, P_y, P_z$ in the free particle. If we wanted to know these values, we would have to perform a measurement on the particle. But a measurement always involves an interaction with the measured object, which breaks the symmetry of the original system by modifying its Hamiltonian (for instance, consider a screen acting as a potential barrier that breaks the homogeneity of space). This means that, under measurement, the particle is no longer free: the symmetry breaking introduced by the interaction with the measuring apparatus is what allows us to have empirical access to an observable that was a symmetry generator of the original free system.

## 5.2 Free particle with spin

The spin $S$ is an internal contribution to the total angular momentum and, therefore, adds further degrees of freedom to the particle: the Hilbert space is now $\mathcal{H} = \mathcal{H}_f \otimes \mathcal{H}_s$, where $\mathcal{H}_f$ is the Hilbert space of the free particle and $\mathcal{H}_s$ is the Hilbert space of the spin. In this case, the Hamiltonian is

$$H = \frac{P^2}{2m} + E_0 \qquad (5\text{-}2)$$

where $E_0$ can only be a multiple of $S^2$ and, then, may be conceived as an internal contribution to the energy (see Ballentine, 1989).

According to our interpretation, in this case the system is composite, because it can be decomposed into two non-interacting subsystems (see IP2): a free particle without spin, represented in $\mathcal{H}_f$ and with Hamiltonian $H_f = P^2/2m$, and a spin system, represented in $\mathcal{H}_s$ and with Hamiltonian $H_s = k S^2$, with $k = const$. Then, the Actualization Rule has to be applied independently to each elemental subsystem.

The rule applies to the free particle subsystem as explained in the previous subsection. On the other hand, in the spin subsystem $H_s$ is invariant under space-rotation: the generators of this symmetry are the three components $J_x, J_y, J_z$ of the total angular momentum $J$. But since in this case the orbital angular momentum $L$ is zero, the total angular momentum $J = L + S$ turns out to be simply $J = S$, and the three components $S_x, S_y, S_z$ of the spin $S$ are the generators of the space-rotation symmetry. Analogously to the case of the free particle, according to our Actualization Rule, in this case $H_s$ acquires a definite value, and also $S^2$ since it is proportional to $H_s$ ($S^2$ is the Casimir operator of the group generated by $S_x, S_y, S_z$); nevertheless, $S_x, S_y, S_z$ are not definite-valued since they are the generators of the symmetry of the Hamiltonian $H_s = k S^2$.

Again, this conclusion agrees with the fact that we have no empirical access to the spin components of the free particle with spin. If we want to know the value of those components, we have to perform a measurement on the system: we have to introduce a magnetic field $B$ of modulus $|B|$ in some direction, say $z$, which breaks the isotropy of space and, as a consequence, the original space-rotation symmetry. Under the action of $B$, the Hamiltonian $H_s$ is not invariant under space-rotation anymore, because now it includes the interaction $-\gamma |B| S_z$ that privileges a particular direction of space. In other words, we can have experimental access to the spin component $S_z$ only by means of a measurement that breaks the space-rotation symmetry of the original Hamiltonian and, therefore, makes the system no longer free. This is the usual way in which a spin component is measured in a Stern-Gerlach experiment (the experiment will be described in detail in Subsection 6.3).

## 5.3 Harmonic oscillator

By definition, a harmonic oscillator is a system with a quadratic potential energy, which produces a restoring force against any displacement from equilibrium that is proportional to the displacement. In one dimension, the Hamiltonian of this system is

$$H = \frac{P^2}{2m} + \frac{m\Omega^2 Q^2}{2} \tag{5-3}$$

where $Q$ is the position observable and $\Omega$ is the frequency of oscillation. If the dimensionless position and momentum operators, $q = (m\Omega/\hbar)^{1/2} Q$ and $p = (1/m\Omega\hbar)^{1/2} P$, are introduced in eq.(5-3), the Hamiltonian reads

$$H = \frac{1}{2} \hbar \Omega \left( q^2 + p^2 \right) \tag{5-4}$$

In turn, if the observable number of modes $N = a^\dagger a$ is used,

$$N = a^\dagger a = \left( \frac{q - ip}{\sqrt{2}} \right)\left( \frac{q + ip}{\sqrt{2}} \right) \quad \Rightarrow \quad H = \hbar \Omega \left( N + \frac{1}{2} \right) \tag{5-5}$$

As it is well known, in this case the spectra of $H$ and $N$ can be obtained algebraically:

$$H |n\rangle = \omega_n |n\rangle \tag{5-6}$$
$$N |n\rangle = n |n\rangle \tag{5-7}$$

The non-degeneracy of $H$ expresses the fact that it has no symmetries: the CSCO $\{H\}$ defines the basis $\{|n\rangle\}$ of the Hilbert space of the system.

According to our Actualization Rule, since in this case the Hamiltonian is non-degenerate, the preferred CSOP is $\{|n\rangle\langle n|\}$: the definite-valued observables of the system are $H$ and all the observables commuting with $H$. In particular, the number of modes $N$ acquires a definite value because $[H, N] = 0$.

The harmonic oscillator has a central relevance in quantum mechanics because it provides a model for many kinds of vibrating systems. In particular, the electromagnetic field can be decomposed in terms of linearly independent modes, each one of which behaves as an harmonic oscillator that is usually associated to a particle; in this case, $N$ is conceived as the observable number of particles. But it is worth emphasizing that the particle-picture of the system, with a definite number of particles, is not generic: it can only be metaphorically retained when the observable $N$ meets the requirements imposed by the Actualization Rule (we shall return on this point in the discussion of ontological matters, in particular, in Subsection 8.6).

Another point to stress here is that, in all of those vibrating phenomena, the energy of the system (and its functions, as $N$) is the relevant physical magnitude of the system, whose value is assumed to be definite. Our interpretation agrees with this usual assumption, since it selects the non-degenerate Hamiltonian as the observable that defines the basis of actualization.

## 5.4 Free hydrogen atom

In the physical literature, the hydrogen atom is described as a two-particles system consisting of an electron and a proton interacting to each other by means of a Coulombian interaction. In this case, the Hamiltonian reads

$$H = \frac{P_e^2}{2m_e} + \frac{P_p^2}{2m_p} + \frac{e^2}{|Q_e - Q_p|} \tag{5-8}$$

where the subindexes $e$ and $p$ refer to the electron and the proton respectively, and $e$ is the electric charge of the electron. The usual strategy for solving the energy eigenvalue equation in coordinate representation is to refer the Hamiltonian to the center of mass of the system by means of a canonical transformation, and to write the resulting equation in spherical coordinates $(r, \theta, \phi)$. As it is well known, with this strategy the solution of the equation can be expressed as the product of two functions, one only dependent on the radial coordinate and the other only dependent on the angular coordinates: $\Psi(r, \theta, \phi) = R(r) Y(\theta, \phi)$. By solving the radial and the angular equations, three "good" quantum numbers are obtained: the principal quantum number $n$, the orbital angular momentum quantum number $l$ and the magnetic quantum number $m_l$. These quantum numbers correspond to the eigenvalues of the observables $H$, $L^2$ and $L_z$ respectively, where $L$ is the orbital angular momentum, and $L_x$, $L_y$, $L_z$ are its components:

$$H |n,l,m_l\rangle = \omega_n |n,l,m_l\rangle \tag{5-9}$$
$$L^2 |n,l,m_l\rangle = l(l+1)\hbar^2 |n,l,m_l\rangle \tag{5-10}$$
$$L_z |n,l,m_l\rangle = m_l \hbar |n,l,m_l\rangle \tag{5-11}$$

with $n = 0,1,2,\ldots$, $l < n$, and $-l \leq m_l \leq l$. In particular, the energy eigenvalues are computed as

$$\omega_n = -\frac{\mu e^4}{2\hbar^2 n^2} \tag{5-12}$$

where $\mu = (1/m_e + 1/m_p)^{-1}$ is the reduced mass of the atom. Therefore, the hydrogen atom is described in terms of the basis $\{|n,l,m_l\rangle\}$ defined by the CSCO $\{H, L^2, L_z\}$: the quantum numbers $n$, $l$, and $m_l$ label the solutions $\Psi_{nlm_l}$ of the energy eigenvalue equation.

In this case, the Hamiltonian is degenerate due to its space-rotation invariance. When the spin of the electron is not considered (for the effect of the spin, see below, Subsection 5.6), the total angular momentum $J = L + S$ is simply $J = L$. Then, the three components $L_x$, $L_y$, $L_z$ of $L$ are the generators of the symmetry group, and $L^2$ is the Casimir operator of the group. As a consequence, although $l$, and $m_l$ are good quantum numbers in the sense of collaborating in the definition of a basis of the Hilbert space, the eigenvalues $\omega_n$ of the Hamiltonian do not depend on them: due to the symmetry of $H$, the values of $L^2$ and $L_z$ have no manifestations in the energy spectrum. According to our Actualization Rule, as the result of the degeneracy of $H$, the observables $L^2$ and $L_z$ do not acquire definite values: the only definite-valued observables of the system are $H$ and the observables having, at least, the same space-rotation symmetry (at least, the same degeneracy) as $H$.

The fact that our interpretation does not confer definite values to $L^2$ and $L_z$ should agree with experimental evidence, in particular, with the data coming from spectroscopy. Let us consider each observable in detail:

a) In quantum chemistry, the states $\Psi_{nlm_l}$ of the atom (orbitals) are labeled as $X\alpha$, where $X$ is the principal quantum number $n$, and $\alpha$ is replaced with $s, p, d, f$, etc., that is, with letters corresponding to the value of the angular momentum quantum number $l$: $1s$: $2s$, $2p$, $3s$, $3p$, $3d$, etc. As we can see, the magnetic quantum number $m_l$ is not included in those labels because, although $\Psi_{nlm_l}$ depends on the three quantum numbers, the space-rotation symmetry of the Hamiltonian makes the selection of $L_z$ a completely arbitrary decision: since space is isotropic, we can choose $L_x$ or $L_y$ to obtain an equally legitimate description of the free atom. The arbitrariness in the selection of the $z$-direction is manifested in spectroscopy by the fact that the spectral lines give no experimental evidence about the values of $L_z$: we have no empirical access to the number $m_l$. Our interpretation, that does not assign a definite value to $L_z$, agrees with those experimental results. Analogously to the case of the free particle with spin (Subsection 5.2), if we want to know the value of $L_z$, we have to introduce a magnetic field that breaks the isotropy of space (we shall describe this situation in detail in the next subsection).

b) On the contrary, the value of the quantum number $l$ is included in the traditional orbitals' labels as $s$, $p$, $d$, etc. Moreover, the value of $l$ can be inferred from the observed energy spectrum of the hydrogen atom, and it plays a role in the explanation of the well-known spectral series (Paschen, Balmer, Lyman, etc.). These facts might be interpreted as a symptom of the definite-valuedness of $L^2$ in the free hydrogen atom. However, the manifestation of the value of $l$ requires the interaction between the atom and an electromagnetic field. The usual explanation runs as follows. Since energy transitions involve the absorption or emission of a photon (spin 1), conservation of the angular momentum forces the atom to experience a change of 1 in its orbital angular momentum $L$. For this reason, when a photon is absorbed by an atom in an $s$ orbital, the atom acquires orbital momentum and makes a transition to a $p$ orbital; when absorbed by an atom in a $p$ orbital, the orbital momentum increases ($p \rightarrow d$ transition) or decreases ($p \rightarrow s$ transition), depending on the relative orientations of the photon and the atom angular momenta. But transitions $s \rightarrow d$ or $p \rightarrow f$ are forbidden. From this explanation, it is clear that the manifestation of the value of $l$ is the result of an interaction; but, then, the system is not the free hydrogen atom anymore. The new system has a Hamiltonian of the form

$$H = H_{at} + H_{em} + H_{int} \qquad (5\text{-}13)$$

where $H_{at}$ is the Hamiltonian of the free hydrogen atom (see eq.(5.8)), and $H_{em}$ is the Hamiltonian of the electromagnetic field, which can be computed as the infinite sum of the Hamiltonians of the independent harmonic oscillators corresponding to the infinite modes of the field (see eq.(5.5)). In turn, $H_{int}$ is the interaction Hamiltonian, that depends on the dipole moment of the atom and on the electric field (see Ballentine, 1989, pp. 548-549). The interaction breaks the original symmetry in $L^2$ and, as a consequence, removes the energy degeneracy in the quantum number $l$: now the energy eigenvalues $\omega_{nl}$ turn out to be functions of both the quantum numbers $n$ and $l$. This fact is what leads to the manifestation of the value of $l$ in the energy spectrum, and allows $L^2$ to become a definite-valued observable in the new, non-free system.

The fact that $L^2$ is not a definite-valued observable in the free *hydrogen* atom does not mean that it never acquires a definite value in a free atom. The particular features of the hydrogen atom strongly depend on the Coulombian potential, conceived as generated by its one-proton nucleus. In more complex atoms, the potential in not perfectly Coulombian, and this asymmetry removes the degeneracy in $l$ of the Hamiltonian: the energy eigenvalues $\omega_{nl}$ are functions of both $n$ and $l$ with no need of interaction (see Ballentine, 1989, p. 280). This means that $L^2$ does no longer discriminate among the different eigenvectors corresponding to a single degenerate energy eigenvalue, but rather removes the degeneracy of the symmetric Coulombian case. According to our Actualization Rule, this implies that $L^2$ is a definite-valued observable for free atoms with non-Coulombian potential.

## 5.5 Zeeman effect

As it is well known, when an external magnetic field is applied to the atom, the spectral lines split into multiple closely spaced lines. First observed by Pieter Zeeman in 1896, this phenomenon is known as Zeeman effect.

In the previous subsection we have seen that, either in the Coulombian or in the non-Coulombian potential case, the Hamiltonian is endowed with a space-rotation symmetry that makes the energy eigenvalues to be independent of the magnetic quantum number $m_l$, that is, to be degenerate in $m_l$. It is precisely due to this symmetry that the selection of $L_z$ for completing the basis of the Hilbert space is the result of an arbitrary decision. The arbitrariness of choosing the $z$-direction agrees with the fact that there is no experimental evidence about the value of $m_l$ in the energy spectrum.

Analogously to the measurement on a free particle with spin (Subsection 5.2), a magnetic field $B$ along the $z$-axis breaks the isotropy of space and, as a consequence, the space-rotation symmetry of the Hamiltonian. In this case, the breaking of the symmetry removes the energy degeneracy in $m_l$: now $L_z$ is not arbitrarily chosen but selected by the direction of the magnetic field. But, in turn, this implies that the atom is no longer free: the Hamiltonian of the new system is approximately (see Ballentine, 1989, p. 326)

$$H = H_{at} + \frac{e}{2m_e e} BL \qquad (5\text{-}14)$$

where, again, $H_{at}$ is the Hamiltonian of the free atom. As a consequence, the original degeneracy of the $(2l+1)$-fold multiplet of fixed $n$ and $l$ is now removed: the energy levels turn out to be displaced by an amount

$$\Delta \omega_{nlm_l} = \frac{e\hbar |B|}{2m_e c} m_l \qquad (5\text{-}15)$$

This means that the Hamiltonian, with eigenvalues $\omega_{nlm_l}$, is now non-degenerate: it constitutes by itself the CSCO $\{H\}$ that defines the preferred basis $\{|n,l,m_l\rangle\}$. According to our Actualization Rule, in this case $H$ and all the observables commuting with $H$ are definite-valued: since this is the case for $L^2$ and $L_z$, in the physical conditions leading to the Zeeman effect both observables acquire definite values.

## 5.6 Fine structure

When the spectral lines of the hydrogen atom corresponding to $n>1$ are examined at a very high resolution, they are found to be closely spaced doublets. This splitting was one of the first experimental evidences of the electron spin. This phenomenon is usually explained by saying that the energy levels of the atom are affected by the "coupling" between the electron spin $S$ and the orbital angular momentum $L$. Now the Hamiltonian of the system reads

$$H = H_{at} + H_s + H_{s-o} \qquad (5\text{-}16)$$

where $H_{at}$ is again the Hamiltonian of the free atom, $H_s = k\,S^2$ is the Hamiltonian of the spin, and $H_{s-o}$ is the Hamiltonian representing the spin-orbit interaction, function of the product $L \cdot S$.

When the spin-orbit interaction is neglected ($H_{s-o} = 0$), the system is composite (see Subsection 5.2) and can be described in terms of the basis $\{|n,l,m_l,s,m_s\rangle = |n,l,m_l\rangle \otimes |s,m_s\rangle\}$, where the $s(s+1)\hbar^2$ are the eigenvalues of $S^2$, and the $m_s\hbar$ are the eigenvalues of $S_z$. But when the spin-orbit interaction is taken into account, the observables $L_z$ and $S_z$ no longer commute with $H$ and, therefore, they are not constants of motion of the system: it is usually said that $m_l$ and $m_s$ are not good quantum numbers anymore. Nevertheless, the Hamiltonian is still invariant under space-rotation: the components $J_x$, $J_y$, $J_z$ of the total angular momentum $J$ are the generators of the symmetry group, and $J^2$ is the Casimir operator of the group, with eigenvalues $j(j+1)\hbar^2$. In turn, $J$ is the sum of the orbital angular momentum $L$ and the spin angular momentum $S$:

$$J = L + S \qquad m_j = m_l + m_s \qquad (5\text{-}17)$$

where $m_j$ corresponds to the eigenvalue of $J_z$. So, now $m_j$ is a good quantum number. But we also know that

$$J^2 = (L+S)^2 \quad \Rightarrow \quad L \cdot S = \frac{J^2 - L^2 - S^2}{2} \qquad (5\text{-}18)$$

This means that $H_{s-o}$ is a function of $J^2$, $L^2$ and $S^2$, and the corresponding quantum numbers $j$, $l$ and $s$ are also good quantum numbers. As a consequence, the eigenvalues of the total Hamiltonian have the general form

$$\omega_{nljs} = \omega_{nl} + \xi(nl)\left[j(j+1) - l(l+1) - s(s+1)\right] \qquad (5\text{-}19)$$

where the $\omega_{nl}$ represent the energy eigenvalues with no spin-orbit coupling, and $\xi$ is a function of $nl$ (see Tinkham, 1964, pp. 181-183). Then, the basis $\{|n,l,j,s,m_j\rangle\}$ of the Hilbert space of the system is defined by the CSCO $\{H, L^2, J^2, S^2, J_z\}$.

It is quite clear that the spin-orbit coupling removes the original degeneracy of the eigenvalues $\omega_{nl}$ of the atom with no coupling. Therefore, in this case our Actualization Rule selects $L^2$, $J^2$ and $S^2$ as definite-valued observables, because all of them commute with $H$ and have the same degeneracy in $m_j$ as $H$. But the space-rotation symmetry still present in the system leads to a degeneracy of $H$, manifested by the fact that the energy eigenvalues $\omega_{nljs}$ do not depend on $m_j$. Then, according to our Actualization Rule, although in this case $m_j$ is a good quantum number, $J_z$ does not acquire a definite value, and this result agrees with the arbitrariness of the selection of the $z$-direction for $J_z$.

When a magnetic field is applied to the atom, the spectral lines split in different ways. The "normal" Zeeman effect, explained in the previous subsection, is observed in spin 0 states where, obviously, the spin-orbit coupling has no effect. In the states where the spin-orbit coupling is effective, the action of the magnetic field produces a further splitting of the energy levels known as "anomalous" Zeeman effect. Nevertheless, the explanation of the anomalous effect is the same as that of the normal effect: the action of the magnetic field along the $z$-axis breaks the space-rotation symmetry of the Hamiltonian by privileging the $z$-direction, and this leads to the removal of the original degeneracy of the Hamiltonian in the quantum number $m_j$ (instead of in the quantum number $m_l$ as in the normal effect). In this case, our Actualization Rule prescribes that $J_z$ will be also definite-valued.

## 5.7 The Born-Oppenheimer approximation

Our Actualization Rule endows the Hamiltonian of the system with the role of selecting the preferred context and, therefore, the energy of the system always acquires an actual definite value. But this does not mean that the momentum is a definite-valued observable in any case, since it does not always commute with the Hamiltonian. In fact, when a system is affected by a scalar field, its Hamiltonian has the general form

$$H = \frac{P^2}{2m} + V(Q) \quad (5\text{-}20)$$

If the mass of the system is small, the kinetic term prevails over the potential term, and the Hamiltonian approximately commutes with $P^2$. In turn, for very large masses, the kinetic term can be neglected and $H$ approximately commutes with $V(Q)$. So, the Actualization Rule supports the usual claim that "small" systems approximately actualize in momentum and "large" systems approximately actualize in position. In this sense, our interpretation agrees with the physical assumption that electrons have definite momentum but not definite position, and the nucleus has definite position but not definite momentum. In general, our rule explains the fact that macroscopic systems, with their large masses, –approximately– posses a definite value of position.

This point has a particular relevance in molecular chemistry, where the description of molecules is based on the adiabatic separation of electron and nuclear motions. As it is well known, the Born-Oppenheimer approximation conceives the nuclei as classical-like particles, that is, as precisely localized objects. This approximation strategy of holding the nucleus at rest in a definite position can be thought off as formally arising from making the masses of the nuclei infinite. However, from a strictly quantum-mechanical viewpoint, without a rule for selecting the definite-valued observables of the system, the assumption of infinite nuclear masses does not explain yet why the nucleus can be treated as having a definite value of position. As Primas says, "we hardly understand why the Born-Oppenheimer *picture* is compatible with the *concepts* of quantum mechanics" (Primas, 1983, p. 13; see also Woolley, 1978; Amann, 1992).

Our interpretation provides an answer to this conceptual problem. For large masses, the Hamiltonian is –approximately– invariant under boost transformation and, therefore, it approximately commutes with position. As a consequence, according to the Actualization Rule, the position observable acquires a definite value: this provides a conceptual justification to the Born-Oppenheimer assumption. Of course, masses are never infinite: this is what makes the Born-Oppenheimer strategy an approximation and not a precise method. But also in this sense our interpretation agrees with the usual assumption: since the Hamiltonian perfectly commutes with position only in the infinite mass limit, only in this limit we can say with absolute precision that position acquires an actual definite value. In real situations, the definite-valued observable will generally be an observable very "similar" to position, but which becomes indistinguishable from position for increasing masses.

The problem of justifying the Born-Oppenheimer approximation is a particular case of the so-called "problem of the classical limit of quantum mechanics", which consists in explaining how the classical description of a macroscopic system may arise from quantum mechanics. As we shall see in Section 7, under certain definite conditions, in macroscopic systems for which $\hbar/S \to 0$, where $S$ is the characteristic action of the system (a more precise limit than the limit $m \to \infty$ of the Born-Oppenheimer approximation), the definite-valued observables manifest themselves as classical-like magnitudes in the classical description.

## 6. The quantum measurement problem

In the standard von Neumann model, a quantum measurement is conceived as an interaction between a system $S$ and a measuring apparatus $M$. Before the interaction, $M$ is prepared in a ready-to-measure state $|r_0\rangle$, eigenvector of the pointer observable $R$ of $M$, and the state of $S$ is a superposition of the eigenstates $|a_i\rangle$ of an observable $A$ of $S$. The interaction introduces a correlation between the eigenstates $|a_i\rangle$ of $A$ and the eigenstates $|r_i\rangle$ of $R$:

$$|\psi_0\rangle = \sum_i c_i |a_i\rangle \otimes |r_0\rangle \quad \rightarrow \quad |\psi\rangle = \sum_i c_i |a_i\rangle \otimes |r_i\rangle \qquad (6\text{-}1)$$

The problem consists in explaining why, being the state $|\psi\rangle$ a superposition of the $|a_i\rangle \otimes |r_i\rangle$, the pointer $R$ acquires a definite value.

In the orthodox collapse interpretation, the pure state $|\psi\rangle$ is assumed to "collapse" to a mixture $\rho^c$:

$$\rho^c = \sum_i |c_i|^2 |a_i\rangle \otimes |r_i\rangle \langle a_i| \otimes \langle r_i| \qquad (6\text{-}2)$$

where the probabilities $|c_i|^2$ are given an ignorance interpretation. Then, in this situation it is supposed that the measuring apparatus is in one of the eigenvectors $|r_i\rangle$ of $R$, say $|r_k\rangle$, and therefore $R$ acquires a definite value $r_k$, the eigenvalue corresponding to the eigenvector $|r_k\rangle$, with probability $|c_k|^2$. In the modal interpretations, the problem is to explain the definite reading of the pointer with its associated probability, without assuming the collapse hypothesis. In our case, the Actualization Rule is what must accomplish this task.

We shall begin our argument by framing the von Neumann model in the context of the measurement practices. In fact, due to the probabilistic nature of quantum mechanics, the maximum information about a quantum system is always obtained by means of repeated measurements on the same system or on identical systems. Therefore, it is necessary to distinguish among:

- *Single measurement*: It is a single process, in which the reading of the pointer is registered. A single measurement, considered in isolation, does not supply yet relevant information about the state of the system $S$.

- *Frequency measurement*: It is a repetition of identical single measurements, whose purpose is to obtain the values $|c_i|^2$ on the basis of the frequencies of the pointer readings in the different single measurements. A frequency measurement supplies relevant information about the state of $S$, but is not yet sufficient to completely identify such a state.

- *State measurement*: It is a collection of frequency measurements, each one of them with its particular experimental arrangement. Each arrangement correlates the pointer $R$ of the apparatus $M$ with an observable $A_i$ of the system, in such a way that the $A_i$ are not only different, but also non-commuting to each other. The information obtained by means of such a collection of frequency measurements is sufficient to reconstruct the state of $S$ (we shall return on this point in Subsection 6.6).

The von Neumann model addresses the quantum measurement problem in the framework of the single measurement. This is completely reasonable to the extent that, if we do not have an adequate explanation of the single case, we cannot account for the results obtained by the repetition of single cases. Nevertheless, although we shall mainly analyze the single measurement case, we shall not forget that a single measurement is always an element of a measurement procedure by means of which, finally, frequencies are to be obtained.

## 6.1 Ideal measurement

In the von Neumann model and, in general, in the discussions about the quantum measurement problem, the Hamiltonians involved in the process are usually not taken into account. In our interpretation, where the Hamiltonians plays a central role, we have to provide a more detailed model of the measurement process. Thus, we shall say that a single measurement is a three-stage process:

- *Stage I* ($t \leq 0$): The system $S$ and the apparatus $M$ do not interact.
- *Stage II* ($0 < t < t_1$): During this stage, $S$ and $M$ interact, and the interaction establishes the correlation.
- *Stage III* ($t \geq t_1$): The interaction ends at $t = t_1$.

**Stage I:** Let us suppose that the single measurement is an element of a state measurement whose purpose is to obtain the coefficients of the state of the elemental quantum system $S: (\mathcal{O}_S \subseteq \mathcal{H}_S \otimes \mathcal{H}_S, H_S)$. Let us also consider a non-degenerate observable $A \in \mathcal{O}_S$:

$$A|a_i\rangle = a_i|a_i\rangle, \quad \text{where } \{|a_i\rangle\} \text{ is a basis of } \mathcal{H}_S \tag{6-3}$$

At a time $t_0 = 0$, the state of $S$ reads[14]

$$|\psi_S(t_0 = 0)\rangle = \sum_i c_i |a_i\rangle \in \mathcal{H}_S \qquad (6\text{-}4)$$

For simplicity, we shall assume that the Hamiltonian $H_S \in \mathcal{O}_S$ of $S$ is non-degenerate:[15]

$$H_S |\omega_{Si}\rangle = \omega_{Si} |\omega_{Si}\rangle, \quad \text{where } \{|\omega_{Si}\rangle\} \text{ is a basis of } \mathcal{H}_S \qquad (6\text{-}5)$$

The measuring apparatus is an elemental quantum system $M: (\mathcal{O}_M \subseteq \mathcal{H}_M \otimes \mathcal{H}_M, H_M)$ having an observable $R \in \mathcal{O}_M$, which has to possess different and macroscopically distinguishable eigenvalues in order to play the role of the pointer:

$$R|r_i\rangle = r_i |r_i\rangle, \quad \text{where } \{|r_i\rangle\} \text{ is a basis of } \mathcal{H}_M \qquad (6\text{-}6)$$

At time $t_0 = 0$, the apparatus $M$ is prepared in a ready-to-measure state $|r_0\rangle$, eigenvector of $R$:

$$|\psi_M(t_0)\rangle = |r_0\rangle \in \mathcal{H}_M \qquad (6\text{-}7)$$

For simplicity, we shall assume that the Hamiltonian $H_M \in \mathcal{O}_M$ of $M$ is non-degenerate:[16]

$$H_M |\omega_{Mi}\rangle = \omega_{Mi} |\omega_{Mi}\rangle, \quad \text{where } \{|\omega_{Mi}\rangle\} \text{ is a basis of } \mathcal{H}_M \qquad (6\text{-}8)$$

For the reading of the pointer to be possible, the eigenvectors $|r_i\rangle$ of $R$ have to be stationary. Thus, the apparatus $M$ is constructed in such a way that $R$ commutes with $H_M$:

$$[H_M, R] = 0 \quad \Rightarrow \quad |\omega_{Mi}\rangle = |r_i\rangle \quad \Rightarrow \quad H_M |r_i\rangle = \omega_{Mi} |r_i\rangle \qquad (6\text{-}9)$$

Then,

$$|\psi_M(t_0 = 0)\rangle = |r_0\rangle = |\psi_M(t_0 + \Delta t)\rangle = |\psi_M\rangle \in \mathcal{H}_M \qquad (6\text{-}10)$$

Therefore, according to the System Composition postulate IP3, at time $t_0 = 0$ the state of the composite system $S \cup M : (\mathcal{O} \subseteq \mathcal{H} \otimes \mathcal{H}, H)$ will be

$$|\psi_I(t_0 = 0)\rangle = |\psi_S(t_0 = 0)\rangle \otimes |\psi_M\rangle = \sum_i c_i |a_i\rangle \otimes |r_0\rangle \in \mathcal{O} \qquad (6\text{-}11)$$

where $\mathcal{O} = \mathcal{O}_S \otimes \mathcal{O}_M$ and $\mathcal{H} = \mathcal{H}_S \otimes \mathcal{H}_M$. Since during Stage I there is no interaction between $S$ and $M$, then $H_{\text{int}} = 0$ and the total Hamiltonian of $S \cup M$ is

$$H = H_S \otimes I_M + I_S \otimes H_M \in \mathcal{O} \tag{6-12}$$

---

[14] If the state of $S$ is prepared at a time $t < t_0 = 0$, it will evolve up to $t_0 = 0$ under the action of the Hamiltonian $H_S$; then, we shall not measure the prepared state but the state at $t_0 = 0$ resulting from the evolution. If we want to measure the state of $S$ as originally prepared, the Hamiltonian $H_S$ must be zero or a multiple of the identity (see Note 11) and, therefore, completely degenerate.

[15] From the complete explanation of the measurement process, it will turn out to be clear that the degeneracy of $H_S$ does not modify the final result (see Note 17).

[16] We shall discuss this assumption in Subsection 6.5, where the conditions for a single measurement will be presented in detail.

According to the System Decomposition postulate IP2, this means that $S$ and $M$ are subsystems of the composite system $S \cup M$.

**Stage II:** In this second, interaction stage, the systems $S$ and $M$ interact through an interaction Hamiltonian $H_{int}$. This means that the composite system $S \cup M : (\mathcal{O} \subseteq \mathcal{H} \otimes \mathcal{H}, H)$ becomes the system $S_{II} : (\mathcal{O} \subseteq \mathcal{H} \otimes \mathcal{H}, H_{II})$, whose Hamiltonian reads

$$H_{II} = H_S \otimes I_M + I_S \otimes H_M + H_{int} = H + H_{int} \in \mathcal{O} \tag{6-13}$$

In turn, the state $|\psi_I(t_0 = 0)\rangle$ of $S \cup M$ in Stage I turns out to be the initial state $|\psi_{II}(t_0 = 0)\rangle$ of $S_{II}$ in Stage II. According to the Dynamical Postulate QP4, such a state evolves to a state $|\psi_{II}(t_1)\rangle$ after a $\Delta t = t_1$:

$$|\psi_{II}(t_1)\rangle = e^{-iH_{II}t_1/\hbar}|\psi_{II}(t_0 = 0)\rangle = e^{-iH_{II}t_1/\hbar}|\psi_I(t_0 = 0)\rangle \in \mathcal{H} \tag{6-14}$$

It can be proved that, if the interaction Hamiltonian $H_{int}$ is

$$H_{int} = -\frac{\lambda \hbar}{t_1}\left(A \otimes P^R\right) \tag{6-15}$$

where $\lambda$ is a constant and $P^R$ is the observable conjugate to $R$, $\left[R, P^R\right] = i\hbar$, then the final state of $S_{II}$ in Stage II is (see Mittelstaedt, 1998)

$$|\psi_{II}(t_1)\rangle = \sum_i c_i |a_i\rangle \otimes |r_i\rangle \in \mathcal{H} \tag{6-16}$$

**Stage III:** At time $t = t_1$ the interaction ends: the system $S_{II}$ becomes the original composite system $S \cup M$, whose Hamiltonian is again $H = H_S \otimes I_M + I_S \otimes H_M \in \mathcal{O}$ (see eq.(6-12)). Since in this stage $H_{int} = 0$, according to the System Decomposition Postulate IP2, $S$ and $M$ turn out to be again subsystems of the composite system $S \cup M$. In turn, the state $|\psi_{II}(t_1)\rangle$ of $S_{II}$ in Stage II becomes the initial state $|\psi_{III}(t_1)\rangle$ of $S \cup M$ in Stage III:

$$|\psi_{III}(t_1)\rangle = |\psi_{II}(t_1)\rangle \in \mathcal{H} \tag{6-17}$$

Since in this stage $S$ and $M$ are quantum systems (they evolve unitarily under their corresponding Hamiltonian) and they also are elemental, we can apply the Actualization Rule to each one of them:

(a) The system $S : (\mathcal{O}_S \subseteq \mathcal{H}_S \otimes \mathcal{H}_S, H_S)$, with initial state $\rho_S(t_1) = Tr_{(M)}\rho_{III}(t_1) = Tr_{(M)}|\psi_{III}(t_1)\rangle\langle\psi_{III}(t_1)|$, evolves unitarily under the action of $H_S$ according to the Dynamical Postulate QP4. However, the preferred CSOP $\{|\omega_{Si}\rangle\langle\omega_{Si}|\}$ is time-invariant since it is defined by the eigenbasis of the –by assumption– non-degenerate $H_S$.[17] Here we have to distinguish two cases:

(a.1) if $[H_S, A] = 0$, then $|\omega_{Si}\rangle = |a_i\rangle$. This means that both $H_S$ and $A$ are definite-valued.

(a.2) if $[H_S, A] \neq 0$, then $|\omega_{Si}\rangle \neq |a_i\rangle$. As a consequence, the observable $A$ is not definite-valued.

(b) The system $M: (\mathcal{O}_M \subseteq \mathcal{H}_M \otimes \mathcal{H}_M, H_M)$, with initial state $\rho_M(t_1) = Tr_{(S)}\rho_{III}(t_1) = Tr_{(S)}|\psi_{III}(t_1)\rangle\langle\psi_{III}(t_1)|$, evolves unitarily under the action of $H_M$ according to the Dynamical Postulate QP4. However, the preferred CSOP is time-invariant since it is defined by the eigenbasis of the non-degenerate $H_M$. In turn, $[H_M, R] = 0$ in order to guarantee the stationarity of the $|r_i\rangle$ (see eq.(6-9)). Therefore, in the apparatus $M$ the preferred CSOP is $\{|\omega_{Mi}\rangle\langle\omega_{Mi}| = |r_i\rangle\langle r_i|\}$, and this means that both $H_M$ and $R$ are definite-valued.

Due to the perfect correlation between the $|r_i\rangle$ and the $|a_i\rangle$, the ideal quantum measurement is usually interpreted as if the reading of a particular value $r_\Omega$ of the apparatus' pointer $R$ were the univocal indication of the corresponding value $a_\Omega$ of the system's observable $A$. However, according to our interpretation this is not the case in situation (a.2), where $A$ is not a definite-valued observable. This fact, which may sound bizarre, turns out to be a non-problematic result when the difference between classical measurement and quantum measurement is clearly understood. The traditional interpretation of the ideal quantum measurement is modeled under the paradigm of classical measurements, based on the correlation between the actual values of an apparatus' pointer and of an observable to be measured. But in quantum measurements the final goal is not to "discover" the actual value of a system's observable, but *to reconstruct the state of the system* just before the beginning of the measurement process. Therefore, the only relevant fact is the definite reading of the apparatus' pointer: the task consists in explaining how the repetition of single measurements where the pointer is definite-valued allows us to reconstruct the state of the measured system. As we have shown, according to our interpretation, no matter whether the system's observable $A$ acquires a definite value or not, in each single measurement *the apparatus' pointer is always definite-*

---

[17] If $H_S \neq 0$ were degenerate, then the preferred CSOP of $S$ would be $\{P_{Si}\}$, where the $P_{Si}$ are the eigenprojectors of $H_S$. Therefore, the definite-valuedness of $A$ would require that $[H_S, A] = 0$ and that $A$ had, at least, the same degeneracy as $H_S$; then, if $A$, as assumed, is non-degenerate, then it does not acquire a definite value. In the case that $H_S = 0$ (or a multiple of the identity, see Note 11), actualization in $S$ does not occur.

*valued.* Now we shall show that the repetition of single measurements in a frequency measurement provides us with the correct coefficients of the system's state.

**Propensities and actualization:** The preferred CSOP $\{|\omega_{Mi}\rangle\langle\omega_{Mi}| = |r_i\rangle\langle r_i|\}$ of the apparatus $M$ defines the set of possible facts $\mathcal{F}^p$. Now we can compute the measure of the propensity to actualization of each possible fact $\langle\langle F[|r_i\rangle\langle r_i|]\rangle\rangle \in \mathcal{F}^p$ (see eq.(4-4)):

$$p^p_{\rho_M(t_1)}\left(\langle\langle F[|r_i\rangle\langle r_i|]\rangle\rangle\right) = Tr\left(\rho_M(t_1)|r_i\rangle\langle r_i|\right) = \langle r_i|\rho_M(t_1)|r_i\rangle \tag{6-18}$$

where the initial state $\rho_M(t_1)$ of $M$ in Stage III is given by (see eqs.(6-16) and (6-17))

$$\rho_M(t_1) = Tr_{(S)}|\psi_{III}(t_1)\rangle\langle\psi_{III}(t_1)| = \sum_{ij} c_i c_j^* |r_i\rangle\langle r_j| \tag{6-19}$$

Therefore,

$$p^p_{\rho_M(t_1)}\left(\langle\langle F[|r_i\rangle\langle r_i|]\rangle\rangle\right) = |c_i|^2 \tag{6-20}$$

Since each possible fact $\langle\langle F[|r_i\rangle\langle r_i|]\rangle\rangle$ is equivalent to the possible facts $\langle\langle F[H_M:\omega_{Mi}]\rangle\rangle$ and $\langle\langle F[R:r_i]\rangle\rangle$ (see IP6), then,

$$p^p_{\rho_M(t_1)}\left(\langle\langle F[H_M:\omega_{Mi}]\rangle\rangle\right) = |c_i|^2 \tag{6-21}$$

$$p^p_{\rho_M(t_1)}\left(\langle\langle F[R:r_i]\rangle\rangle\right) = |c_i|^2 \tag{6-22}$$

As we can see, the measures of the propensities so obtained agree with the probabilities assigned to the different readings of the pointer by the collapse interpretation. Moreover, those measures turn out to be constant with time, a reasonable result for guaranteeing the repeatability of the single measurement.

Of course, if we want to experimentally obtain the $|c_i|^2$, we have to repeat the single measurement under the same conditions for the propensities to manifest themselves as frequencies in the resulting frequency measurement. But in each single measurement, one and only one possible fact $\langle\langle F[|r_i\rangle\langle r_i|]\rangle\rangle$ will become actual (see IP10), say $\langle\langle F[|r_\Omega\rangle\langle r_\Omega|]\rangle\rangle$. Therefore,

- $F[H_M:\omega_{M\Omega}]$: the case-property $[H_M:\omega_{M\Omega}]$ corresponding to the type-property $[H_M]$ actually occurs (the Hamiltonian $H_M$ actually has the value $\omega_{M\Omega}$).

- $F[R:r_\Omega]$: the case-property $[R:r_\Omega]$ corresponding to the type-property $[R]$ actually occurs (the pointer $R$ actually has the value $r_\Omega$).

Moreover, in the composite system $S \cup M$ (see IP4) the case-property $[I_S \otimes R:r_\Omega]$ corresponding to the type-property $[I_S \otimes R]$ actually occurs, since $[R] = [I_S \otimes R]$. In turn, for the same reason we know that $[H_S \otimes I_M : \omega_{S\Phi}]$ actually occurs in $S \cup M$ when $[H_S : \omega_{S\Phi}]$ actually occurs in $S$, and that $[I_S \otimes H_M : \omega_{M\Omega}]$ actually occurs in $S \cup M$ when $[H_M : \omega_{M\Omega}]$ actually occurs in $M$. Therefore, in the composite system $S \cup M$, where $H = f(H_S \otimes I_M, I_S \otimes H_M) = H_S \otimes I_M + I_S \otimes H_M$, the case-property $[H : (\omega_{S\Phi} + \omega_{M\Omega})]$ corresponding to the type-property $[H]$ also actually occurs: the total Hamiltonian $H$ actually has the value $(\omega_{S\Phi} + \omega_{M\Omega})$. In other words, the total energy of the composite system is the sum of the energies of the component subsystems, as always assumed in practice.

### 6.2 Non-ideal measurement

Two kinds of non-ideal measurements are usually distinguished in the literature:

- *Imperfect measurement* (*first kind*):

$$\sum_i c_i |a_i\rangle \otimes |r_0\rangle \;\rightarrow\; \sum_{ij} d_{ij} |a_i\rangle \otimes |r_j\rangle \quad \text{where, in general, } d_{ij} \neq 0 \text{ with } i \neq j \qquad (6\text{-}23)$$

- *Disturbing measurement* (*second kind*):

$$\sum_i c_i |a_i\rangle \otimes |r_0\rangle \;\rightarrow\; \sum_i c_i |a_i^d\rangle \otimes |r_i\rangle \quad \text{where, in general, } \langle a_i^d | a_j^d \rangle \neq \delta_{ij} \qquad (6\text{-}24)$$

However, the disturbing measurement can also be expressed as an imperfect measurement by a change of basis:

$$\sum_i c_i |a_i^d\rangle \otimes |r_i\rangle = \sum_{ij} d_{ij} |a_i\rangle \otimes |r_j\rangle \qquad (6\text{-}25)$$

In certain modal interpretations (Kochen-Dieks, Vermaas-Dieks), the rule of property-ascription, when applied to non-ideal measurements, leads to results that disagree with those obtained in the orthodox collapse interpretation (see Albert & Loewer, 1990, 1993). If the properties ascribed by a modal interpretation are different from those ascribed by the collapse interpretation, the question is how different they are. In the case of an imperfect measurement, it can be expected that the $d_{ij} \neq 0$, with $i \neq j$, be small; then, the difference might be also small. But in the case of a disturbing measurement, the $d_{ij} \neq 0$, with $i \neq j$, need not be small and, as a consequence, the disagreement between the properties ascribed by the modal interpretation and those ascribed by collapse might be unacceptable (see a full discussion in Bacciagaluppi & Hemmo, 1996). This fact has been considered by Harvey Brown as a "silver bullet" for killing the modal interpretations (cited in Bacciagaluppi & Hemmo, 1996).

We shall not distinguish between the two kinds of non-ideal measurements because the result of the application of our Actualization Rule does not depend on the values of the off-diagonal terms $d_{ij}$. As we shall see, according to our interpretation, the observable $R$ that plays the role of the apparatus' pointer acquires a definite value in any case.

**Stages I to III:** In a non-ideal measurement, Stage I is characterized in the same way as in the ideal case. The difference begins at Stage II, where the correlation introduced by the interaction Hamiltonian $H_{int}$ is not perfect. Therefore, the final state $|\psi_{II}(t_1)\rangle$ of Stage II, which is the initial state $|\psi_{III}(t_1)\rangle$ of Stage III, reads

$$|\psi_{III}(t_1)\rangle = |\psi_{II}(t_1)\rangle = \sum_{ij} d_{ij} |a_i\rangle \otimes |r_j\rangle \qquad (6\text{-}26)$$

Since in Stage III $H_{int} = 0$, according to the System Decomposition postulate IP2, $S$ and $M$ are again subsystems of the composite system $S \cup M$. As we have discussed in the case of the ideal measurement, we are not interested in the definite-valued observables of $S$; so, we shall analyze the result of the process in the apparatus $M$.

The system $M$ begins Stage III in an initial state

$$\rho_M(t_1) = Tr_{(S)} \rho_{III}(t_1) = Tr_{(S)} |\psi_{III}(t_1)\rangle\langle\psi_{III}(t_1)| = \sum_n \langle a_n | \psi_{III}(t_1)\rangle\langle\psi_{III}(t_1) | a_n \rangle \qquad (6\text{-}27)$$

Then,

$$\rho_M(t_1) = \sum_{ij} \rho_{Mij} |r_i\rangle\langle r_j| \qquad (6\text{-}28)$$

where

$$\rho_{Mij} = \sum_n d_{ni} d_{nj}^* \qquad (6\text{-}29)$$

Although $M$ evolves unitarily under the action of $H_M$ according to the Dynamical Postulate QP4, the preferred CSOP is time-invariant since it is defined by the eigenbasis of $H_M$. In turn, since $H_M$ commutes with $R$, the preferred CSOP is again $\{|\omega_{Mi}\rangle\langle\omega_{Mi}| = |r_i\rangle\langle r_i|\}$, and this means that both $H_M$ and $R$ are definite-valued.

**Propensities and actualization:** In this case, the time-invariant measure of the propensity to actualization of each possible fact $\langle\langle F[|r_i\rangle\langle r_i|]\rangle\rangle \in \mathcal{F}^p$ is given by

$$p_{\rho_M(t_1)}^p \left(\langle\langle F[|r_i\rangle\langle r_i|]\rangle\rangle\right) = \langle r_i | \rho_M(t_1) | r_i \rangle = \rho_{Mii} = \sum_n |d_{ni}|^2 = |d_{ii}|^2 + \sum_{n \neq i} |d_{ni}|^2 \qquad (6\text{-}30)$$

As we can see, if the coefficients $d_{ni}$, with $n \neq i$, of the off-diagonal terms of the initial state in Stage III are zero (see eqs.(6-23) or (6-26)), we are in the ideal measurement case, where $\rho_{Mii} = |d_{ii}|^2 = |c_i|^2$. If the coefficients $d_{ni}$, with $n \neq i$, are not zero, we are in the non-ideal measurement case. However, in this case two situations have to be distinguished:

- If the $d_{ni}$, with $n \neq i$, are small in the sense that $\sum_{n \neq i} |d_{ni}|^2 \ll |d_{ii}|^2$ (see eq.(6-30)), then $\rho_{Mii} \simeq |d_{ii}|^2 \simeq |c_i|^2$. This means that, in the frequency measurement performed by repetition of this single measurement, the coefficients $|c_i|^2$ can be *approximately obtained*.

- If the $d_{ni}$, with $n \neq i$, are not small, then $\rho_{Mii} \simeq |d_{ii}|^2$ does not hold. Therefore, the result obtained by means of the frequency measurement will be *non reliable*.

Nevertheless, no matter whether the result of the frequency measurement is reliable or not, in each single measurement one and only one possible fact $\langle\langle F[|r_i\rangle\langle r_i|]\rangle\rangle$ will become actual, say $\langle\langle F[|r_\Omega\rangle\langle r_\Omega|]\rangle\rangle$. Therefore, we can guarantee that, in the measuring apparatus $M$,

$$F[H_M : \omega_{M\Omega}] \quad \text{and} \quad F[R : r_\Omega] \tag{6-31}$$

Moreover, if $F[H_S : \omega_{S\Phi}]$ is an actual fact in the system $S$, then in the composite system $S \cup M$, $F[H : (\omega_{S\Phi} + \omega_{M\Omega})]$ and $F[I_S \otimes R : r_\Omega]$ (see the last paragraph of Subsection 6.1).

Summing up, Albert and Loewer (1990, 1993) are right in claiming that the ideal measurement is a situation that can never be achieved in practice: the interaction in Stage II never introduces a completely perfect correlation; in spite of this, physicists usually perform successful measurements. Our account of the measurement process clearly shows that perfect correlation is not a necessary condition for "good" measurements: the coefficients of the system's state at the beginning of the process can be approximately obtained even when the correlation is not perfect, if the reliability condition of small off-diagonal terms is satisfied. Nevertheless, both in the reliable and in the non reliable frequency measurement, in each single measurement *a definite reading of the pointer $R$ is obtained*: our interpretation is immune to Brown's "silver bullet".

## 6.3 The Stern-Gerlach experiment

Since the Stern-Gerlach experiment is the paradigm of quantum measurement, it is worth while to see how all the elements of our general account of measurement can be found in this case.

The experiment is usually described as follows. A neutral free particle with spin,[18] with constant velocity in the $y$-direction, passes between the poles of a magnet that produces an inhomogeneous magnetic field $B$, with components $B_x = B_y = 0$ and $B_z = zB'$, where $B'$ is the field gradient. The particle is described in the plane $zy$, and in a frame of reference moving uniformly in the $y$-direction, where $P_y = 0$. The gradient of the magnetic field produces a force that deflects the particle in the $z$-direction: the deflection depends on the component of spin in that direction.

---

[18] The deflection of a charged particle by the Lorentz force would obscure the spin dependence of the deflection.

As we have seen in Subsection 5.2, the free particle with spin is a composite system $S_s \cup S_f$. In this measurement situation:

➢ the spin subsystem $S_s$, represented in $\mathcal{H}_s$ and with Hamiltonian $H_s = k S^2$, is the system under measurement $S$.

➢ the free particle without spin $S_f$, represented in $\mathcal{H}_f$ and with Hamiltonian $H_f = P_z^2 / 2m$, has to be a part of a measuring apparatus $M$ such that $[H_M, P_z] = 0$: this guarantees that the eigenvectors of $P_z$ are stationary and, then, $P_z$ can play the role of the pointer (see the discussion of this point in Subsection 6.5).

On this basis, at Stage I we find that:

- The observable $A$ is the spin in $z$-direction, $S_z \in \mathcal{O}_S \subseteq \mathcal{H}_S \otimes \mathcal{H}_S$:
$$S_z |\uparrow\rangle = s_\uparrow |\uparrow\rangle \quad , \quad S_z |\downarrow\rangle = s_\downarrow |\downarrow\rangle \qquad (6\text{-}32)$$
where $s_\uparrow = -s_\downarrow = (1/2)\hbar$.

- The momentum in $z$-direction plays the role of the pointer, $P_z \in \mathcal{O}_M \subseteq \mathcal{H}_M \otimes \mathcal{H}_M$:
$$P_z |+\rangle = p_+ |+\rangle \quad , \quad P_z |-\rangle = p_- |-\rangle \quad , \quad P_z |0\rangle = p_0 |0\rangle \qquad (6\text{-}33)$$
where $\{|+\rangle, |-\rangle, |0\rangle\}$ is a basis of $\mathcal{H}_M$.[19]

- The states of $S$ and $M$ are, respectively, $|\psi_S\rangle = c_1 |\uparrow\rangle + c_2 |\downarrow\rangle$ and $\psi_M = |0\rangle$. Then,
$$|\psi_I(t_0 = 0)\rangle = c_1 |\uparrow\rangle \otimes |0\rangle + c_2 |\downarrow\rangle \otimes |0\rangle \qquad (6\text{-}34)$$

- As we have said, the Hamiltonian of $S$ is $H_S = k S^2$, and the Hamiltonian of $M$ is such that $[H_M, P_z] = 0$. Therefore,
$$H_M |+\rangle = \omega_+ |+\rangle \quad , \quad H_M |-\rangle = \omega_- |-\rangle \quad , \quad H_M |0\rangle = \omega_0 |0\rangle \qquad (6\text{-}35)$$

**Ideal measurement:** At Stage II, the total Hamiltonian $H_{II} = H_M + H_{int}$ introduces a perfect correlation. Then, the initial state of $S \cup M$ in Stage III is
$$|\psi_{III}(t_1)\rangle = c_1 |\uparrow\rangle \otimes |+\rangle + c_2 |\downarrow\rangle \otimes |-\rangle \qquad (6\text{-}36)$$

---

[19] Again, here we are assuming that $\{|+\rangle, |-\rangle, |0\rangle\}$ is a basis of $\mathcal{H}_M$ and that $H_M$ is non-degenerate for simplicity. We shall discuss these assumptions in Subsection 6.5.

The initial state of the subsystem $M$ then reads

$$\rho_M(t_1) = Tr_{(S)} |\psi_{III}(t_1)\rangle \langle \psi_{III}(t_1)| = |c_1|^2 |+\rangle\langle +| + |c_2|^2 |-\rangle\langle -| \qquad (6\text{-}37)$$

The preferred CSOP of $M$ is given by the eigenbasis of the Hamiltonian $H_M$ and, then, it is $\{|+\rangle\langle +|, |-\rangle\langle -|, |0\rangle\langle 0|\}$. Since $[H_M, P_z] = 0$, both the Hamiltonian $H_M$ and the momentum $P_z$ in $z$-direction are definite-valued. The measure of the propensity to actualization of the possible facts corresponding to $P_z$ can be computed as

$$p^p_{\rho_M(t_1)}\left(\langle\langle F[|+\rangle\langle +|]\rangle\rangle\right) = \langle +|\rho_M(t_1)|+\rangle = |c_1|^2 \qquad (6\text{-}38)$$

$$p^p_{\rho_M(t_1)}\left(\langle\langle F[|-\rangle\langle -|]\rangle\rangle\right) = \langle -|\rho_M(t_1)|-\rangle = |c_2|^2 \qquad (6\text{-}39)$$

$$p^p_{\rho_M(t_1)}\left(\langle\langle F[|0\rangle\langle 0|]\rangle\rangle\right) = \langle 0|\rho_M(t_1)|0\rangle = 0 \qquad (6\text{-}40)$$

As expected, these measures are time-invariant: they do not depend on the time when the reading of the point is performed, that is, on the precise position where the detectors are placed in Stage III. If the measures of those propensities depended on the instantaneous state of the system, the result of the frequency measurement would be extremely sensitive to the precise location of the detectors: any imperceptible perturbation would substantially modify the frequencies so obtained, making the frequency measurement physically unrealizable. In turn, since the possible facts with measure zero do not become actual (see IP11), only one of the following two situations will be actual:

$$F[H_M : \omega_+] \quad \text{and} \quad F[P_z : p_+] \qquad (6\text{-}41)$$

$$F[H_M : \omega_-] \quad \text{and} \quad F[P_z : p_-] \qquad (6\text{-}42)$$

**Non-ideal measurement:** In this case, $H_{II}$ does not introduce a perfect correlation. The initial state of $S \cup M$ in Stage III is, then,

$$|\psi_{III}\rangle = d_{11}|\uparrow\rangle \otimes |+\rangle + d_{12}|\uparrow\rangle \otimes |-\rangle + d_{21}|\downarrow\rangle \otimes |+\rangle + d_{22}|\downarrow\rangle \otimes |-\rangle \qquad (6\text{-}43)$$

The initial state of the subsystem $M$ reads

$$\rho_M(t_1) = \rho_{M11}|+\rangle\langle +| + \rho_{M12}|+\rangle\langle -| + \rho_{M21}|-\rangle\langle +| + \rho_{M22}|-\rangle\langle -| \qquad (6\text{-}44)$$

where

$$\rho_{Mij} = \sum_{n=1}^{2} d_{ni} d_{nj}^* \qquad (6\text{-}45)$$

In other words,

$$\rho_M(t_1) = \begin{pmatrix} |d_{11}|^2 + |d_{21}|^2 & d_{11}d_{12}^* + d_{21}d_{22}^* \\ d_{12}d_{11}^* + d_{22}d_{21}^* & |d_{12}|^2 + |d_{22}|^2 \end{pmatrix} \qquad (6\text{-}46)$$

The preferred CSOP of $M$ is again given by the eigenbasis of the Hamiltonian $H_M$ and, since $[H_M, P_z] = 0$, both the Hamiltonian $H_M$ and the momentum $P_z$ in $z$-direction are definite-valued. But now the measure of the propensity to actualization of the corresponding possible facts results

$$p^p_{\rho_M(t_1)}\left(\left\langle\left\langle F[|+\rangle\langle+|]\right\rangle\right\rangle\right) = \langle+|\rho_M(t_1)|+\rangle = |d_{11}|^2 + |d_{21}|^2 \tag{6-47}$$

$$p^p_{\rho_M(t_1)}\left(\left\langle\left\langle F[|-\rangle\langle-|]\right\rangle\right\rangle\right) = \langle-|\rho_M(t_1)|-\rangle = |d_{22}|^2 + |d_{12}|^2 \tag{6-48}$$

$$p^p_{\rho_M(t_1)}\left(\left\langle\left\langle F[|0\rangle\langle 0|]\right\rangle\right\rangle\right) = \langle 0|\rho_M(t_1)|0\rangle = 0 \tag{6-49}$$

In this non-ideal case, the frequency measurement resulting from the repetition of this single measurement will be reliable if $|d_{21}|^2 \ll |d_{11}|^2$ and $|d_{12}|^2 \ll |d_{22}|^2$; if not, the frequency measurement will not supply the necessary information for the reconstruction of the original state of the measured system. Nevertheless, the observable $P_z$ acquires a definite value in any case, and this is the prediction that can be directly tested in each single measurement.

This analysis of the Stern-Gerlach experiment allows us to point out a feature of the quantum measurement that cannot be noticed in the merely formal treatments of the process. In fact, in the von Neumann model, the observable $A$ of the system $S$ under measurement is considered in formal terms and deprived of its physical content. Then, the interaction between $S$ and the measuring apparatus $M$ is endowed with the only role of introducing the correlation between $A$ and the pointer $R$. However, the varied physical situations described in Section 5 show that we have no empirical access to the observables that are generators of the symmetries of the system's Hamiltonian; in the context of measurement, $A$ may be one of those observables. This is precisely the case in the Stern-Gerlach experiment, where $S_z$ is a generator of the space-rotation symmetry of $H_s = k\,S^2$. It is the interaction with the magnetic field $B = B_z$ what breaks the isotropy of space by privileging the $z$-direction and, as a consequence, breaks the space-rotation symmetry of $H_s$ (see Subsection 5.2). This physical account of the measurement shows that, when the observable $A$ is a generator of a symmetry of the Hamiltonian $H_S$ of $S$, the interaction with the apparatus $M$ breaks that symmetry and, at the same time, establishes the correlation between $A$ and $R$. Therefore, from a physical viewpoint, in these cases measurement can be conceived as a process that breaks the symmetries of the system to be measured and, in this way, allows us to reconstruct its state in terms of an otherwise empirically inaccessible symmetry-generator observable.[20]

## 6.4 Infinite tails

An argument that stresses the difficulties introduced by non-ideal measurements is that posed by Elby (1993) in the context of the Stern-Gerlach experiment. This argument points to the fact that the wavefunctions in $z$-variable typically have infinite "tails" that introduce non-zero cross-terms; therefore, the "tail" of the wavefunction of the "down" beam may produce detection in the upper detector, prepared to detect $p_+$, and vice versa.

Let us consider this new argument in detail by supposing that the imperfection is due to a non-perfect collimation of the incoming beam. In this case, with the magnetic field still turned off, we would obtain a diffuse spot instead of a definite point on the screen. Therefore, the perfect ready-to-measure state $|r_0\rangle = |0\rangle$ has to be replaced with a narrow Gaussian $|\varphi_0(z)\rangle$. As a consequence, the measurement process turns out to be expressed as

$$|\psi_I\rangle = \left(c_1|\uparrow\rangle + c_2|\downarrow\rangle\right) \otimes |\varphi_0(z)\rangle \rightarrow |\psi_{III}\rangle = c_1|\uparrow\rangle \otimes |\varphi_+(z)\rangle + c_2|\downarrow\rangle \otimes |\varphi_-(z)\rangle \quad (6\text{-}50)$$

where now $|\varphi_+(z)\rangle$ and $|\varphi_-(z)\rangle$ are Gaussians that do not need to be as narrow as the initial one. Let us call the widths of the upper and the lower detectors $\Delta z_+$ and $\Delta z_-$ respectively. Thus, the long tail of the Gaussian $|\varphi_+(z)\rangle$ arrives to $\Delta z_-$ and the long tail of the Gaussian $|\varphi_-(z)\rangle$ arrives to $\Delta z_-$. We can compute the probabilities corresponding to the four possible cases:

$$p(\uparrow,+) = \left|\left(\langle\uparrow| \otimes \langle\varphi_+(z)|\right)|\psi_{III}\rangle\right|^2 = |c_1|^2 \int_{\Delta z_+} \left\||\varphi_+(z)\rangle\right\|^2 dz = |c_{11}|^2 \quad (6\text{-}51)$$

$$p(\uparrow,-) = \left|\left(\langle\uparrow| \otimes \langle\varphi_-(z)|\right)|\psi_{III}\rangle\right|^2 = |c_1|^2 \int_{\Delta z_-} \left|\langle\varphi_-(z)|\varphi_+(z)\rangle\right|^2 dz = |c_{12}|^2 \quad (6\text{-}52)$$

$$p(\downarrow,+) = \left|\left(\langle\downarrow| \otimes \langle\varphi_+(z)|\right)|\psi_{III}\rangle\right|^2 = |c_2|^2 \int_{\Delta z_+} \left|\langle\varphi_+(z)|\varphi_-(z)\rangle\right|^2 dz = |c_{21}|^2 \quad (6\text{-}53)$$

$$p(\downarrow,-) = \left|\left(\langle\downarrow| \otimes \langle\varphi_-(z)|\right)|\psi_{III}\rangle\right|^2 = |c_2|^2 \int_{\Delta z_-} \left\||\varphi_-(z)\rangle\right\|^2 dz = |c_{22}|^2 \quad (6\text{-}54)$$

where

$$p^P_{\rho_M(t_1)}\left(\langle\langle F[|+\rangle\langle +|]\rangle\rangle\right) = p(\uparrow,+) + p(\downarrow,+) = |c_{11}|^2 + |c_{21}|^2 \quad (6\text{-}55)$$

$$p^P_{\rho_M(t_1)}\left(\langle\langle F[|-\rangle\langle -|]\rangle\rangle\right) = p(\downarrow,-) + p(\uparrow,-) = |c_{22}|^2 + |c_{12}|^2 \quad (6\text{-}56)$$

---

[20] The idea is that the formal von Neumann model of quantum measurement can be complemented by a physical model in terms of which measurement is a symmetry breaking process that renders a symmetry generator of the system's Hamiltonian empirically accessible. Although this idea seems plausible, it should be supported by the analysis of further physical measurement processes; this analysis is beyond the limits of the present paper and will be the object of a future study.

According to Elby's argument, these cases can be read as follows:

* $|c_{11}|^2$ is the probability that $|\uparrow\rangle$ be detected by $\Delta z_+$
* $|c_{12}|^2$ is the probability that $|\uparrow\rangle$ be detected by $\Delta z_-$ (tail)
* $|c_{21}|^2$ is the probability that $|\downarrow\rangle$ be detected by $\Delta z_+$ (tail)
* $|c_{22}|^2$ is the probability that $|\downarrow\rangle$ be detected by $\Delta z_-$

Our interpretation shows that, if the reliability condition $|c_{21}|^2 \ll |c_{11}|^2$ and $|c_{12}|^2 \ll |c_{22}|^2$ holds, then the collimation, even if not perfect, is good enough for measurement, since $|c_{11}|^2 \simeq |c_1|^2$ and $|c_{22}|^2 \simeq |c_2|^2$. If the original Gaussian is not very narrow or the screen is placed too far from the magnet, the measurement will be non-reliable since the $c_{ij}$, with $i \neq j$, are not small enough. Nevertheless, according to the Actualization Rule, since the preferred CSOP is defined by the eigenbasis of $H_M$ and the pointer commutes with $H_M$, we obtain a definite reading of the pointer, that is, a definite detection in $\Delta z_+$ or $\Delta z_-$.

## 6.5 Defining measurement

On the basis of this analysis of the quantum measurement, now we can explicitly formulate the conditions that a quantum process must satisfy to be considered a single measurement:

(i) There must be two quantum systems: a system to measure, $S: (\mathcal{O}_S \subseteq \mathcal{H}_S \otimes \mathcal{H}_S, H_S)$, and a measuring apparatus, $M: (\mathcal{O}_M \subseteq \mathcal{H}_M \otimes \mathcal{H}_M, H_M)$, which do not interact at $t < 0$.

(ii) The apparatus $M$ must be constructed in such a way to have an observable $R \in \mathcal{O}_M$ such that $[H_M, R] = 0$, $R$ has, at least, the same degeneracy as $H_M$, and its eigenvalues are different and macroscopically distinguishable.

(iii) Since $t = 0$ and during a period $\Delta t = t_1$, $S$ and $M$ must interact through an interaction Hamiltonian $H_{\text{int}} \neq 0 \in \mathcal{O}_S \otimes \mathcal{O}_M$. The interaction is intended to introduce a correlation between an observable $A \in \mathcal{O}_S$ of $S$ and the observable $R$ of $M$.

Let us note that we have not included the requirement of perfect correlation as a defining condition of single measurement. In fact, even if the correlation is not perfect, in Stage III the energy of the system $M$ is always definite-valued and, as a consequence, the pointer $R$ always acquires a definite value. However, in spite of the fact that we always obtain a definite reading of the pointer in each single measurement, the frequency measurement resulting from the repetition of single measurements is not always *reliable*. If the interaction Hamiltonian introduces an imperfect correlation in such a way that

$$|\psi_I\rangle = \sum_i c_i |a_i\rangle \otimes |r_0\rangle \quad \rightarrow \quad |\psi_{III}\rangle = \sum_{ij} d_{ij} |a_i\rangle \otimes |r_j\rangle \tag{6-57}$$

then,

- the frequency measurement will be *reliable* only when the $d_{ij}$, with $i \neq j$, are small enough to make $|d_{ii}|^2 \simeq |c_i|^2$.

- when the $d_{ij}$, with $i \neq j$, are not small enough, the frequency measurement will be *non reliable*, because the obtained frequencies do not give us the values $|c_i|^2$.

Nonetheless, in both cases each element of the frequency measurement is legitimately a *single measurement*, where *a definite reading of the pointer is obtained*.

In spite of the fact that in the previous subsections we assumed that the Hamiltonian $H_M$ was non-degenerate, this assumption can be relaxed in such a way that

$$H_M = \sum_i \omega_{Mi} P_{Mi} \tag{6-58}$$

where the eigenprojectors $P_{Mi}$ do not need to be one-dimensional. In this case, the preferred CSOP is $\{P_{Mi}\}$. Then, the definite-valuedness of the pointer requires that $R$ have, at least, the same degeneracy –the same symmetries– as $H_M$:

$$R = \sum_i r_i P_{Mi} \tag{6-59}$$

However, in the effective practice of physics, the apparatus is a macroscopic system, whose Hamiltonian is the result of the interaction among a huge number of degrees of freedom. Since in general symmetries are broken by interactions, the symmetry of the Hamiltonian decreases with the complexity of the system. Then, a macroscopic system having a Hamiltonian with symmetries is a highly exceptional situation: in the generic case, the energy is the only constant of motion of the macroscopic system. As a consequence, in practice $H_M$ is non-degenerate, and $\{|\omega_{Mi}\rangle\}$ is a basis of the Hilbert space $\mathcal{H}_M$ of the apparatus. On the other hand, the pointer $R$ cannot have a so huge number of eigenvalues as $H_M$, because the experimental physicists have to be able to discriminate among them (see, for instance, $P_z$ in the Stern-Gerlach experiment, with its three eigenvalues). This means that, in general, $R$ is a "collective" observable (see Omnés, 1994; 1999), that is, a highly degenerate observable whose eigenprojectors introduce a sort of "coarse-graining" in $\mathcal{H}_M$. Nevertheless, given that $[H_M, R] = 0$, and since $R$ has more degeneracies than $H_M$ (which is non-degenerate), $R$ will be a definite-valued observable. Moreover, the time-invariant measures of the propensities of the possible facts $\langle\langle F[R:r_i]\rangle\rangle$ can be computed in terms of the eigenprojectors $P_{Mi}$ of $R$ (see eq.(6-59)): $p^p_{\rho_M}\left(\langle\langle F[R:r_i]\rangle\rangle\right) = Tr(\rho P_{Mi})$ (see eq.(4-5)).[21]

The discussion of the previous point, in turn, allows us to clarify a matter that still remains rather obscure in the present-day literature on quantum measurement. During the last decades, and under the influence of the works of Zurek and his collaborators (see, for instance, Paz & Zurek, 2002; Zurek, 2003), the claim that the measuring device necessarily interacts with its environment has become a commonplace. However, such a claim, that expresses the main basis of the environment-induced approach to decoherence (we shall return on this point in Section 7), faces the conceptual challenge of supplying a precise definition of quantum system, a challenge that has not been solved in the context of that theoretical framework: "In particular, one issue which has been often taken for granted is looming big, as a foundation of the whole decoherence program. It is the question of what are the 'systems' which play such a crucial role in all the discussions of the emergent classicality. This issue was raised earlier, but the progress to date has been slow at best" (Zurek, 1998, p. 122). We have argued elsewhere (Castagnino, Laura & Lombardi, 2007) that this account of decoherence is misleading, since it hides the fact that a single, unitary evolving, quantum system can be partitioned in many different ways on the basis of the observables considered "relevant" and "irrelevant" in each case (see also Harshman & Wickramasekara, 2007a, b; Omnés, 1999): each partition leads to non-unitarily evolving parts which may or may not decohere. Therefore, decoherence is a phenomenon relative to the space of the relevant observables selected in each particular situation.

In contrast to Zurek's view, our interpretation offers a definition of quantum system that is precise and, at the same time, does not contradict the Dynamical Postulate of quantum mechanics. On the basis of this definition, the measuring apparatus cannot be conceived as a material, macroscopic device surrounded by a "bath" of particles interacting with it. The measuring apparatus is the entire quantum system $M$ that interacts with the measured system $S$: it is that system what has a pointer $R$ commuting with its, in general, non-degenerate Hamiltonian $H_M$. As we have pointed out, in the generic case $M$ is a macroscopic system with a huge number of degrees of freedom, and $R$ must be a "collective" and empirically accessible observable to play the role of the pointer; it is for this reason that the many degrees of freedom corresponding to the degeneracies of $R$ are conceived as an "internal environment" by the environment-interaction approach (see Omnés, 1994; 1999). In turn, the requirement $[H_M, R] = 0$ has a clear physical meaning: it is essential to guarantee the stationary behavior of $R$. If it did not hold because of the uncontrollable interaction between the macroscopic device and an external "bath", the reading of $R$ would constantly change and measurement would be impossible. It is precisely at this point that the skills of the experimental physicist play a central role: he has to be capable of designing an experimental arrangement such that the uncontrollable degrees of freedom of the complete system interacting with the measured system do not affect significantly the stationary character of the pointer. This goal may be achieved by different means: by shielding enough the macroscopic device, by making the Hamiltonian of the device much greater than the sum of the Hamiltonian corresponding to the external environment and the interaction Hamiltonian, etc. But, in any case, the measurement has to be a controlled situation where the behavior of the observable $R$ can be used to obtain meaningful information about the measured system $S$.

---

[21] We are grateful to one of the referees for urging us to discuss the assumption of the non-degeneracy of the apparatus' Hamiltonian.

### 6.6 Ensembles and state measurement

According to the ensemble interpretations of quantum mechanics, the theory does not describe single systems, but ensembles of systems: probability assignments only make sense on ensembles (see Ballentine, 1970). It is quite clear that we do not endorse this kind of interpretations. From our modal-Hamiltonian perspective, the quantum state, either a pure state or a mixture, refers to a single system, and the probabilities introduced by the theory measure the ontological propensity to actualization of the possible facts involving that single system.

However, it is also clear that probabilistic predictions can only be tested by means of frequencies on collection of identical systems, that is, on ensembles: a frequency measurement, whose purpose is to obtain the values $|c_i|^2$ of the measured state, is a repetition of single measurements on an ensemble. In general, the statistical predictions of the theory, as the expectation value of an observable in a given state, can be empirically tested only on ensembles (we shall return on this point in Subsection 7.3).

When the concept of ensemble comes into play, not only the frequency measurement, but also the state measurement can be properly understood. In fact, up to this point we have analyzed single and frequency measurements in the case where the system $S$ is initially in a pure state $|\psi_S\rangle$:

$$|\psi_S\rangle = \sum_i c_i |a_i\rangle \qquad (6\text{-}60)$$

In this case, the frequency measurement, if reliable, will supply –at least, approximately– the values $|c_i|^2$. But the same analysis can be made in the general case, where the initial state of $S$ is expressed as $\rho_S$ (pure state or mixture):

$$\rho_S = \sum_{ij} \rho_{ij} |a_i\rangle\langle a_j| \qquad (6\text{-}61)$$

In this case, the frequency measurement, if reliable, will supply –at least, approximately– the values of the coefficients $\rho_{ii}$ of the main diagonal of $\rho_S$ in the basis $\{|a_i\rangle\}$ (where $\rho_{ii} = |c_i|^2$ if $\rho_S = |\psi_S\rangle\langle\psi_S|$). For this purpose, the observable $A$ of $S$ is used:

$$A = \sum_i a_i |a_i\rangle\langle a_i| \qquad (6\text{-}62)$$

and the apparatus $M$, with a Hamiltonian $H_M$ and a pointer $R$ such that $[H_M, R] = 0$, interacts with the system $S$ through an interaction Hamiltonian $H_{\text{int}}$ (see eq.(6-15)):

$$H_{\text{int}} = -\frac{\lambda \hbar}{t_1} \left( A \otimes P^R \right) \qquad (6\text{-}63)$$

But if we also want to know –at least, approximately– the remaining coefficients $\rho_{ij}$ with $i \neq j$, we have to perform further frequency measurements, where the observables $B_{ij}$ and $C_{ij}$ of $S$ are used (see Ballentine, 1998):

$$B_{ij} = 1/2 \left( |a_i\rangle\langle a_j| + |a_j\rangle\langle a_i| \right) \quad , \quad C_{ij} = 1/2 \left( |a_i\rangle\langle a_j| - |a_j\rangle\langle a_i| \right) \qquad (6\text{-}64)$$

since[22]

$$\langle B_{ij} \rangle_{\rho_S} = Tr(\rho_S B_{ij}) = 1/2 (\rho_{ij} + \rho_{ji}) = \text{Re}(\rho_{ij}) \qquad (6\text{-}65)$$

$$\langle C_{ij} \rangle_{\rho_S} = Tr(\rho_S C_{ij}) = 1/2 (\rho_{ij} - \rho_{ji}) = \text{Im}(\rho_{ij}) \qquad (6\text{-}66)$$

In this case, all the frequency measurements are performed on the same ensemble, that is, on a collection of identical systems in the same state $\rho_S$. However, each frequency measurement requires its own experimental arrangement. In particular, the apparatus $M$ has to be constructed in such a way that it interacts with the system $S$ through interaction Hamiltonians $H_{\text{int}}^{Bij}$ or $H_{\text{int}}^{Cij}$ such that

$$H_{\text{int}}^{Bij} = -\frac{\lambda \hbar}{t_1} \left( B_{ij} \otimes P^R \right) \quad , \quad H_{\text{int}}^{Cij} = -\frac{\lambda \hbar}{t_1} \left( C_{ij} \otimes P^R \right) \qquad (6\text{-}67)$$

---

[22] The computation of the $\rho_{ii}$ can also be viewed under this perspective, considering the observable $A$ of eq.(6-62) as $A = \sum_i a_i A_{ii}$, where $A_{ii} = |a_i\rangle\langle a_i|$. Then, $\langle A_{ii} \rangle_{\rho_S} = Tr(\rho_S A_{ii}) = \langle a_i |\rho_S| a_i\rangle = \rho_{ii}$, namely, a particular case of eq.(6-65).

These interaction Hamiltonians introduce the desired correlation between the eigenstates of the observables $B_{ij}$ and $C_{ij}$ of $S$, respectively, and the eigenstates of the pointer observable $R$ of $M$.

Summing up, if we want to reconstruct the state $\rho_S$ of a system $S$, we have to perform a state measurement on an ensemble of systems identical to $S$. Such a state measurement consists of a collection of frequency measurements, each one of which relies on the correlation between the pointer observable of the apparatus $M$ and a particular observable of $S$, where the observables of $S$ used for this purpose are not only different, but also non-commuting to each other. The information obtained by means of this state measurement is sufficient to reconstruct the state $\rho_S$ of $S$.

## 7. Upstairs: the classical limit of quantum mechanics

In the previous section we have given an account of the quantum measurement, where the definite value of the pointer is explained in terms of the actualization of one of the possible facts defined by the apparatus' preferred context. But a measurement is a controlled situation, where the experimental physicist is able to select the set of possible facts where actualization occurs by manipulating the apparatus $M$, in particular, its Hamiltonian $H_M$. However, in nature certain possible facts become actual with no human intervention. Moreover, in certain situations quantum systems have features that admit a classical description. As we announced at the end of Subsection 5.7, the problem of the classical limit is to explain how and under what conditions a classical description arises from an underlying quantum description.

At present, there is broad agreement about the idea that the classical behavior of quantum systems arises after the process called *decoherence*. According to the orthodox view, known as *environment induced decoherence* (*EID*), decoherence is the result of the interaction of an open quantum system with its environment, which selects the candidates to classical states (Paz & Zurek, 2002; Zurek, 2003). In previous works we have presented a different approach, called *self-induced decoherence* (*SID*), according to which a closed quantum system may decohere under well-defined conditions (Castagnino & Laura, 2000; Castagnino & Ordóñez, 2004; Castagnino & Lombardi, 2004, 2005a). On this basis, we have developed a precise account of the classical limit of quantum mechanics, which shows that, if the system is macroscopic enough, after self-induced decoherence it can be described as a classical-statistical system (Castagnino & Lombardi, 2003, 2005b; Castagnino, 2005; Castagnino & Gadella, 2006).

In this section we shall recall the main results of the SID approach and of our account of the classical limit, in order to give them a new reading in the light of the modal-Hamiltonian interpretation just introduced.

### 7.1 Self-induced decoherence

The SID approach is not based on the C*-algebraic framework; rather, the starting point of its formalism is a nuclear algebra $\mathcal{A}$, in terms of which the space of observables $\mathcal{O}$ is defined.

Let us consider a quantum system $S:(\mathcal{O} \subset \mathcal{A}, H)$ with initial state $\rho_0 \in \mathcal{O}'$. Here we are interested in the classical limit, that is, the classical behavior manifested by macroscopic systems. In turn, we have stressed that, due to the huge number of degrees of freedom of a macroscopic system, the non-degeneracy of its Hamiltonian is highly generic (see Subsection 6.5). Therefore, here we shall consider the case of a non-degenerate Hamiltonian with no loss of generality. If $H$ has a continuous spectrum, it reads

$$H = \int_0^\infty \omega |\omega\rangle\langle\omega| d\omega \qquad (7\text{-}1)$$

In the eigenbasis $\{|\omega\rangle\}$ of $H$, any observable $O$ belonging to the van Hove space $\mathcal{V}_O^{VH} \subset \mathcal{O}$ (see van Hove, 1955; Castagnino & Lombardi, 2004) can be expressed as

$$O = \int_0^\infty O(\omega)|\omega) d\omega + \int_0^\infty \int_0^\infty O(\omega,\omega')|\omega,\omega') d\omega d\omega' \qquad (7\text{-}2)$$

where $|\omega) = |\omega\rangle\langle\omega|$ and $|\omega,\omega') = |\omega\rangle\langle\omega'|$ are the generalized eigenvectors of the observable $O$, $\{|\omega),|\omega,\omega')\}$ is a basis of $\mathcal{V}_O^{VH}$, $O(\omega)$ is a generic distribution and $O(\omega,\omega')$ is a regular function. In turn, the initial state $\rho_0$ is a linear functional belonging to $\mathcal{V}_S^{VH}$, the dual space of $\mathcal{V}_O^{VH}$, and it can be expressed as

$$\rho_0 = \int_0^\infty \rho(\omega)(\omega| d\omega + \int_0^\infty \int_0^\infty \rho(\omega,\omega')(\omega,\omega'| d\omega d\omega' \qquad (7\text{-}3)$$

where $\{(\omega|,(\omega,\omega'|\}$ is the basis of $\mathcal{V}_S^{VH}$ (that is, the cobasis of $\{|\omega),|\omega,\omega')\}$), and $\rho(\omega)$ and $\rho(\omega,\omega')$ are regular functions. Thus, the expectation value of any observable $O \in \mathcal{V}_O^{VH}$ in a state $\rho(t) = e^{iHt/\hbar} \rho_0 e^{-iHt/\hbar} \in \mathcal{V}_S^{VH}$ is computed as

$$\langle O \rangle_{\rho(t)} = \int_0^\infty \rho(\omega) O(\omega) d\omega + \int_0^\infty \int_0^\infty \rho(\omega,\omega') O(\omega,\omega') e^{-i(\omega-\omega')t/\hbar} d\omega d\omega' \qquad (7\text{-}4)$$

Under these conditions, we can apply the Riemann-Lebesgue theorem according to which, for $t \to \infty$, the second term of the right-hand side of eq.(7-4) vanishes; thus,

$$\lim_{t \to \infty} \langle O \rangle_{\rho(t)} = \int_0^\infty \rho(\omega)\, O(\omega)\, d\omega = \langle O \rangle_{\rho_*} \qquad (7\text{-}5)$$

where

$$\rho_* = \int_0^\infty \rho(\omega)\, (\omega|\, d\omega \qquad (7\text{-}6)$$

Summing up, according to this approach, self-induced decoherence is a process that leads the expectation value of any observable $O \in \mathcal{V}_O^{VH}$ to a value that can be computed as if the system were in a state represented by the "diagonal" functional $\rho_*$:[23]

$$\lim_{t \to \infty} \langle O \rangle_{\rho(t)} = \langle O \rangle_{\rho_*} \qquad (7\text{-}7)$$

Strictly speaking, the SID approach applies when the Hamiltonian $H$ has continuous spectrum. Nevertheless, it is easy to prove that approximate decoherence can be obtained under conditions of quasi-continuity, that is, in discrete models where (i) the energy spectrum is quasi-continuous, i.e., has a small discrete energy spacing, and (ii) the functions of energy used in the formalism are such that the sums in which they are involved can be approximated by Riemann integrals (see, for instance, Casati & Chirikov, 1995a, 1995b; Gaioli, García Alvarez & Guevara, 1997).

## 7.2 The classical limit

According to our account of the classical limit, the classical description of a quantum system requires two elements: decoherence, as explained by the SID approach, and macroscopicity, meaning that the characteristic action $S$ of the system is much greater than $\hbar$. Then, the task is to represent the diagonal functional $\rho_*$ resulting from decoherence in the corresponding phase space by means of the Wigner transformation (see Hillery, O'Connell, Scully & Wigner, 1984), and to apply the macroscopic limit $\hbar \to 0$ (strictly speaking, $\hbar/S \to 0$):

$$\rho^c(\phi) = \lim_{\hbar \to 0} W \rho_* \qquad (7\text{-}8)$$

where $\phi = (\boldsymbol{q}, \boldsymbol{p}) = (q_1, \cdots, q_{N+1}, p_1, \cdots, p_{N+1})$ is a point of the phase space $\Gamma = \mathbb{R}^{2(N+1)}$ and $W$ is the Wigner transformation, whose definition has to be adequately extended to be applicable to singular distributions. When this task is undertaken, it can be proved that the function $\rho^c(\phi)$ has the following form (see Castagnino, 2005; Castagnino & Gadella, 2006):

$$\rho^c(\phi) = \int_\omega \rho(\omega)\, \delta(H(\phi) - \omega)\, d\omega \qquad (7\text{-}9)$$

where $H(\phi)$ is the Wigner transformation of the Hamiltonian $H$, $H(\phi) = \mathrm{W}H$. In other words, $\rho^c(\phi)$ is an infinite sum of classical densities $\delta(H(\phi) - \omega)$, averaged by the corresponding value of the function $\rho(\omega)$. In turn, $\rho(\omega)$ is a normalized and non-negatively defined function due to its origin, since it represents the diagonal elements of the functional $\rho_0$; this fact is what permits it to be interpreted as a probability function. Therefore, the classical distribution $\rho^c(\phi)$ defined on the phase space $\Gamma = \mathbb{R}^{2(N+1)}$ can be conceived as an infinite sum of classical densities, defined by the global constant of motion $H(\phi) = \omega$, and weighted by their corresponding probabilities $\rho(\omega)$ given by the initial state $\rho_0$.[24] This means that the Hamiltonian turns out to be a classical global constant of motion in the description resulting from the classical limit. Moreover, the basis where the time-invariant functional $\rho_*$ becomes diagonal corresponds precisely to the preferred CSOP $\{|\omega\rangle\langle\omega|\}$ where actualization occurs.

---

[23] The SID approach provides a method for computing the decoherence time $t_D$, that is, the relaxation time of the decoherence process. In particular, $t_D$ can be obtained in terms of the poles of the analytical continuation of the resolvent of $H$ (see Castagnino & Lombardi, 2005a). In a free evolving system, whose Hamiltonian has no poles, $t_D \to \infty$ and decoherence is only nominal.

Once $\rho^c(\phi)$ has been obtained, we can explain the classical limit in the phase space language. As we have seen, self-induced decoherence implies the convergence of the expectation value of any observable $O \in \mathcal{V}_O^{\mathrm{VH}}$ to a final value $\langle O \rangle_{\rho_*}$ (see eq.(7-5)). In turn, we know that the Wigner transformation has the property of preserving expectation values:

$$\langle O \rangle_{\rho(t)} = \langle O(\phi) \rangle_{\rho(\phi,t)} \qquad \langle O \rangle_{\rho_*} = \langle O(\phi) \rangle_{\rho_*(\phi)} \qquad (7\text{-}10)$$

where $O(\phi) = \mathrm{W}O$, $\rho(\phi,t) = \mathrm{W}\rho(t)$ and $\rho_*(\phi) = \mathrm{W}\rho_*$. But we also know that the limit $\hbar \to 0$ turns $\mathrm{W}\rho_*$ into $\rho^c(\phi)$ (see eq.(7-8)). Therefore, in the macroscopic limit, the expectation value of any observable $O \in \mathcal{V}_O^{\mathrm{VH}}$ converges to a final value that can be computed in classical terms as

$$\lim_{t\to\infty}\langle O \rangle_{\rho(t)} = \lim_{t\to\infty}\langle O(\phi) \rangle_{\rho(\phi,t)} = \langle O(\phi) \rangle_{\rho^c(\phi)} \qquad (7\text{-}11)$$

where

$$\langle O(\phi) \rangle_{\rho^c(\phi)} = \int_\omega \rho(\omega)\, O(\omega)\, \delta(H(\phi) - \omega)\, d\omega \qquad (7\text{-}12)$$

This means that the state $\rho(\phi,t)$ *weakly* converges to $\rho^c(\phi)$:

$$w - \lim_{t\to\infty} \rho(\phi,t) = \rho^c(\phi) = \int_\omega \rho(\omega)\, \delta(H(\phi) - \omega)\, d\omega \qquad (7\text{-}13)$$

---

[24] Since here the dimension of the Wigner phase space is $2(N+1)$ and the classical description has the energy as the only global constant of motion, the system is *non-integrable*, as expected due to its macroscopic nature. As we have showed in previous works (Castagnino & Lombardi, 2006, 2007), if $g$ is the number of global constants of motion, in the non-integrable case $(N+1) - g$ local constants of motion can be defined, which permit the characterization of the lower levels of the quantum ergodic hierarchy.

## 7.3 The ontological interpretation of the classical limit

When we explained the measurement process, we distinguished between single measurement and frequency measurement. In the first case, the actualization of one of the possible facts $\langle\langle F[|r_i\rangle\langle r_i|]\rangle\rangle \in \mathcal{F}^p$ defined by the preferred CSOP $\{|\omega_{Mi}\rangle\langle\omega_{Mi}| = |r_i\rangle\langle r_i|\}$ is what accounts for the definite value of the pointer. But for propensities to manifest themselves as frequencies it is necessary to perform a frequency measurement by repeating the single measurement under the same conditions: only in this way can the probabilistic predictions of the theory be tested. In the case of the classical limit, the situation is analogous. Each single elemental quantum system satisfies the Actualization Rule and, as a consequence, one possible fact $\langle\langle F[|\omega\rangle\langle\omega|]\rangle\rangle \in \mathcal{F}^p$ defined by the preferred CSOP $\{|\omega\rangle\langle\omega|\}$ becomes actual; therefore, $H$ and all the observables commuting with $H$ acquire their definite values. However, only when we work with an *ensemble* of quantum systems, statistical predictions can be empirically tested. When these considerations are taken into account, the classical limit can be endowed with a precise ontological meaning in the light of the modal-Hamiltonian interpretation.

Let us consider an elemental quantum system $S:(\mathcal{O}\subset\mathcal{A},H)$, with initial state $\rho_0 \in \mathcal{O}'$, such that its classical limit is explained as in the previous subsection. If we now consider an ensemble of quantum systems identical to $S$, *in each member of the ensemble*:

- At time $t = t_0$, each possible fact $\langle\langle F[|\omega\rangle\langle\omega|]\rangle\rangle$ has a propensity to actualization whose measure is given by the diagonal terms of $\rho_0$ expressed in the preferred basis $\{|\omega\rangle\}$: $p_\rho^p(\langle\langle F[|\omega\rangle\langle\omega|]\rangle\rangle) = \rho(\omega)$ (see IP7).
- Since the preferred CSOP is $\{|\omega\rangle\langle\omega|\}$, one and only one of the possible facts $\langle\langle F[|\omega\rangle\langle\omega|]\rangle\rangle$, say $\langle\langle F[|\omega_\Omega\rangle\langle\omega_\Omega|]\rangle\rangle$, is also an actual fact $F[|\omega_\Omega\rangle\langle\omega_\Omega|]$ (see IP10).
- If $F[|\omega_\Omega\rangle\langle\omega_\Omega|]$ is an actual fact, $H$ and all the observables $R_I$ commuting with $H$ indeterministically acquire definite values $\omega_\Omega$ and $r_{I\Omega}$, respectively: $F[H:\omega_\Omega]$, $F[R_I:r_{I\Omega}]$.

Since this holds for each member of the ensemble, the propensity whose measure is given by $\rho(\omega)$ manifests itself as a *frequency of actual facts in the ensemble*.

Now we can endow the expectation values involved in the classical limit with a precise ontological interpretation. For this purpose we have to distinguish between two cases: the case of the observables that acquire definite values and the case of the remaining observables.

a) The definite-valued observables are diagonal in the energy basis, corresponding to the preferred CSOP. This means that their expectation values are time-independent (see eq.(7-4)):

$$\langle H \rangle_{\rho(t)} = \int_0^\infty \rho(\omega)\, \omega\, d\omega = \langle H \rangle_{\rho_*} = \langle H(\phi) \rangle_{\rho_*(\phi)} \tag{7-14}$$

$$\langle R_I \rangle_{\rho(t)} = \int_0^\infty \rho(\omega)\, r_I\, d\omega = \langle R_I \rangle_{\rho_*} = \langle R_I(\phi) \rangle_{\rho_*(\phi)} \tag{7-15}$$

and, in the macroscopic limit $\hbar \to 0$, they can be computed as (see eq.(7-12))

$$\langle H \rangle_{\rho(t)} = \langle H(\phi) \rangle_{\rho^c(\phi)} = \int_\omega \rho(\omega)\, \omega\, \delta(H(\phi)-\omega)\, d\omega \tag{7-16}$$

$$\langle R_I \rangle_{\rho(t)} = \langle R_I(\phi) \rangle_{\rho^c(\phi)} = \int_\omega \rho(\omega)\, r_I\, \delta(H(\phi)-\omega)\, d\omega \tag{7-17}$$

In each system of the ensemble, $H$ and the $R_I$ indeterministically acquire definite values $\omega_\Omega$ and $r_{I\Omega}$ with a propensity measured by $\rho(\omega_\Omega)$. Therefore, the time-independent expectation values $\langle H \rangle_{\rho(t)} = \langle H(\phi) \rangle_{\rho^c(\phi)}$ and $\langle R_I \rangle_{\rho(t)} = \langle R_I(\phi) \rangle_{\rho^c(\phi)}$ measure the expectation values corresponding to the definite values actually acquired by the observables $H$ and $R_I$ in the members of the ensemble, expectation values that can be computed in terms of the frequencies in the ensemble.

b) Let us now consider any observable $A$ non commuting with $H$. According to our interpretation, it does not acquire a definite value in this system. Nevertheless, since each possible fact $\langle\langle F[A:a_i]\rangle\rangle$ corresponding to $A$ has its propensity to actualization, whose measure depends on the system's state (see IP7), the expectation value $\langle A \rangle_\rho$ can be computed (and even tested when the state is measured by means of a state measurement, see Subsection 6.6). And since the state, that codifies the propensities, changes with time, such an expectation value also changes with time. However, the classical limit shows that, if the system is macroscopic enough, after decoherence $\langle A \rangle_{\rho(t)}$ settles down in a constant value that can be computed as $\langle A(\phi) \rangle_{\rho^c(\phi)}$, where $A(\phi) = WA$ and $\rho^c(\phi)$ is a classical distribution (see eqs.(7-12) and (7-13)). This means that, in the classical limit and *from the viewpoint of the expectation values*, the quantum ensemble can be described as a classical-statistical ensemble in spite of the fact that, in each system of the quantum ensemble, the observable $A$ *does not acquire a definite value*.

This peculiar feature of the quantum ensembles allows us to understand other statistical properties described by quantum mechanics. In particular, according to our interpretation, the uncertainty principle does not refer to single systems but to quantum ensembles. When we say that $[A,B] = iC$ ($C \neq 0$) and, therefore, $\Delta A\, \Delta B \geq 1/2 |\langle C \rangle|$ (see Ballentine, 1998, pp. 223-224), both $\Delta A$ and $\Delta B$ can be computed in terms of the respective expectation values $\langle A \rangle_\rho$ and $\langle B \rangle_\rho$ in the quantum ensemble. However, the dispersions $\Delta A$ and $\Delta B$ have no meaning in each single system since $A$ or $B$, but not both, may be definite-valued.

Moreover, in the classical limit both expectation values can be computed as the expectation values of the classical observables $A(\phi) = WA$ and $B(\phi) = WB$ in a classical-statistical ensemble described by $\rho^c(\phi)$ (see eq.(7-12)). This is what allows us to treat the quantum ensemble with the theoretical tools of classical-statistical mechanics: we can imagine that we are working with a classical-statistical ensemble where the classical observables $A(\phi)$ and $B(\phi)$ have definite values in each classical member of the ensemble, with dispersions $\Delta A$ and $\Delta B$ in the ensemble, respectively.[25] But *this classical description* is valid only when the theory is applied to an ensemble and, as a consequence, it *cannot be used to draw ontological conclusions about single systems*.

Summing up, the fact that our interpretation conceives quantum mechanics as describing single systems does not imply to ignore the statistical results referring to ensembles supplied by the theory. But those statistical results are not taken as the starting point of the interpretation, as in the case of the ensemble interpretations. On the contrary, their meaning is explained in terms of the ontological interpretation of the basic elements of the theory, which are referred to single quantum systems.

## 8. Downstairs: the structure of the ontology

In general, the discussions about the modal interpretations of quantum mechanics are concerned with the traditional conceptual problems of the theory; for instance, one of the main purposes is to solve the measurement problem overcoming the challenges posed by the no-go theorems. Of course, this task is hard enough to concentrate the attention of the participants in the discussion. But these preoccupations should not lead us to forget certain relevant ontological issues, in particular, the questions about the nature of the items referred to by quantum mechanics.

In this section we shall explicitly address the philosophical question about the structure of the ontology referred to by quantum mechanics, in particular, about the *basic categories* of such an ontology. Roughly speaking, we want to know which kinds of items can be assumed to exist as described by the theory, without contradicting its formal structure and its physical content.

---

[25] Of course, if we want to test these predictions, we have to perform different frequency measurements on the ensemble, each one of which will measure one observable with the adequate interaction Hamiltonian, as explained in Subsection 6.6.

## 8.1 Possibilities

The nature of possibility has been one of the most controversial issues in the history of philosophy. However, two general conceptions can be identified, both of which find their roots in Antiquity. One of them, which is usually called "*actualism*", is the conception that reduces possibility to actuality. This was the position of Diodorus Cronus; in Cicero's words, "Diodorus defines the possible as that which either is or will be" (cited in Kneale & Kneale, 1962, p. 117). This view survived over the centuries up to our time; for instance, for Bertrand Russell "possible" means "sometimes", whereas "necessary" means "always" (Russell, 1919). The other conception, called "*possibilism*", conceives possibility as an ontologically irreducible feature of reality. From this perspective, the stoic Crissipus defined possible as "that which is not prevented by anything from happening even if it does not happen" (cited in Bunge, 1977, p. 172). In present day metaphysics, the debate actualism-possibilism is still alive. For the actualists, the adjective "actual" is redundant: non-actual possible items (objects, properties, facts, etc.) do not exist, they are nothing. According to the possibilists, on the contrary, not every possible item is an actual item: possible items –*possibilia*– constitute a basic ontological category (see Menzel, 2007).

As we have seen, according to modal interpretations, the formalism of quantum mechanics does not determine what actually is the case, but rather describes what *may* be the case, that is, possible facts with their corresponding probabilities. Once the definite-valued observables (type-properties) are selected by a certain rule of property-ascription, the actual occurrence of a particular value of such observables (a case-property) is an essentially indeterministic phenomenon which, as a consequence, cannot be determined by the theory. This means that, for each definite-valued observable, among all the possibilities described by the theory, only one is actually realized: the remaining possibilities do not become actual, and they might never become actual in the particular system under consideration. Nonetheless, from the realist perspective underlying modal interpretations, if quantum mechanics were true, it would describe reality. So, which is the reality accounted for by the theory? Certainly, not actual reality: if quantum mechanics is about what may be the case, it describes *possible reality*.

On this basis, in our interpretation of quantum mechanics ontological propensities embody a possibilist, non-actualist possibility: a possible fact does not need to become actual to be real. This possibility is defined by the postulates of quantum mechanics and is not reducible to actuality. This means that reality spreads out in two realms, the *realm of possibility* and the *realm of actuality*. In Aristotelian terms, being can be said in different ways: as possible being or as actual being. And none of them is reducible to the other.

The items populating *the realm of possibility* belong to different ontological categories: type-properties with their corresponding case-properties, possible facts and propensities. In a particular elemental quantum system (see the table at the end of Section 4):

- The *case-properties* $[O{:}o_i]$ of each *type-property* $[O]$ are the properties that may occur.
- The *possible facts* $\langle\langle F[O{:}o_i] \rangle\rangle$ are the possible occurrence of the case-properties $[O{:}o_i]$.
- The *propensities to actualization* measured by $p_\rho^\alpha(\langle\langle F[O{:}o_i]\rangle\rangle)$, since applied to the possible facts $\langle\langle F[O{:}o_i]\rangle\rangle$, are second-order type-properties and their corresponding values are second-order case-properties.

Many authors reject an ontology of non-actual *possibilia* claiming that there is no non-trivial criterion of identity for non-actual items (see, e.g. Quine, 1953). In order to face this challenge, possibilists have developed different strategies directed to the identification of *possibilia*: possible worlds, subsistence as different than existence, etc. Vulcano, as the innermost planet between Sun and Mercury, or Julius Caesar having a sixth finger in his right hand are the kind of possible objects considered in the discussions. It is clear that these are not the cases involved in our conception of possibility: we are not proposing an all-embracing theory of *possibilia* that should be applied to any modal sentence. Our possibilist conception only applies to possibility in quantum mechanics, where the possible facts are clearly defined by the structure of the quantum system. In other words, the criterion of identity for possible items is given by the theory, which also fixes the space of admissible possibilities. In fact, the *ontological structure of the realm of possibility* is embodied in the definition of the elemental quantum system $S: (\mathcal{O} \subseteq \mathcal{H} \otimes \mathcal{H}, H)$, with its initial state $\rho_0 \in \mathcal{O}'$:

- The *space of observables* $\mathcal{O}$ identifies:
  - All the type-properties with their corresponding case-properties.
  - All the possible facts and the equivalence relationships among them.
  - All the Boolean sets of possible facts.

- The *initial state* $\rho_0$ codifies the measures of the propensities to actualization of all the possible facts at the initial time. These propensities evolve deterministically according to the Schrödinger equation.

From this perspective, the realm of possibility is not less real than the realm of actuality: a possible fact exists as possible even if it never becomes actual. Propensities are real second-order properties that follow a deterministic evolution independently of which possible facts become actual.

The fact that propensities belong to the realm of possibility does not mean that they do not have physical consequences in the realm of actuality. On the contrary, propensities produce definite effects on actual reality even if they never become actual. An interesting manifestation of such effectiveness is the case of the so-called "non-interacting experiments" (Elitzur & Vaidman, 1993; Vaidman, 1994), where non-actualized possibilities can be used in practice, for instance, to test bombs without exploding them (Penrose, 1994). This shows that possibility is a way in which reality manifests itself, a way independent of and not less real than actuality.

We might ask ourselves what logical framework would be the formalism adequate to express this notion of possibility. As it is well known, traditional systems of modal logic are *extensions* of the classical propositional logic: they are based on the classical calculus but extend it by adding modal operators with their corresponding inference rules (Haack, 1974, 1978). But the set of quantum propositions does not have the Boolean structure of classical propositional logic and, of course, the addition of modal operators does not cancel that non-Boolean character and the resulting contextual nature of quantum mechanics (Domenech, Freytes & de Ronde, 2006). Our interpretation favors an ontological strategy to face this difficulty. As we have seen, each propensity measured by $p_\rho^\alpha$ applies to its corresponding set $\mathcal{F}^\alpha$ of possible facts defined by the CSOP $\{\Pi_\alpha\}$; then, given two different CSOP's $\{\Pi_\alpha\}$ and $\{\Pi_\beta\}$, $p_\rho^\alpha$ and $p_\rho^\beta$ measure different propensities. From an ontological viewpoint, this means that $p_\rho^\alpha$ and $p_\rho^\beta$ represent the measure of *different second-order type-properties*, whose corresponding case-properties −values− are assigned to the possible facts belonging to two different sets, $\mathcal{F}^\alpha$ and $\mathcal{F}^\beta$ respectively. Therefore, if we want to introduce modalities to quantum propositions, it seems reasonable to do it in each context: each set $\mathcal{F}^\alpha$ of possible facts will have its own modal operators to be applied to the corresponding propositions; in particular, the modal operator of possibility of the context defined by the CSOP $\{\Pi_\alpha\}$ will express the physical possibility −the propensity− measured by $p_\rho^\alpha$. But since all the sets of possible facts preserve their Boolean structure, the traditional systems of modal logic can be used for this purpose. Of course, this move restrains us from making modal assertions with propositions coming from non-commuting contexts. Nonetheless, this does not imply a limitation to generality, because we know the context where actualization occurs since the very beginning. From our perspective, in the modal proposition "*It is possible that the observable $A$ has the value $a_1$ or the observable $B$ has the value $b_1$*", where $A$ and $B$ do not commute, the use of the term '*possible*' is ambiguous since it refers to different second-order properties, that is, to the different propensities of the possible facts corresponding to $[A: a_1]$ and $[B: b_1]$. As a consequence, that modal proposition cannot even be meaningfully formulated, and this is perfectly consistent with the fact that there is no single measurement by means of which a truth value may be assigned to that proposition. In other words, since any quantum system univocally defines its own preferred context, we do not need to extend modalities beyond the limits of each context.

Summing up, our interpretation provides an ontological picture where the realm of possibility is as real as the realm of actuality: it is populated by properties, possible facts and propensities related to each other in a well-defined structure. It is precisely this feature what endows our modal interpretation with a *modal* character that is not merely semantical or epistemic, but mainly *ontological*.

## 8.2 Probabilities

It is usually said that our modern conception of probability has had a dual nature since its emergence in the mid-seventeenth century (see Hacking, 1975). In its *epistemic* sense, it measures the extent to which evidence supports a given hypothesis. In its *ontological* sense, it describes regularities exhibited in nature.

In the twentieth century, two widely recognized analyses of epistemic probability have been proposed, one called "logical" −the tradition of Keynes (1921) and Carnap (1950)−, and the other called "subjective" −represented by Ramsey (1926) and de Finetti (1974). In turn, ontological probability was originally identified with limiting relative frequencies in infinite sequences −e.g. Reichenbach (1949) and von Mises (1957). But since the 1950's, the idea of interpreting ontological probabilities as propensities −Popper (1957, 1959)− began to be considered by a number of philosophers as an alternative to the frequentist interpretation.

As we have seen, in our interpretation of quantum mechanics probability measures the ontological propensity to actualization of a possible fact. Therefore, it is not defined by epistemic notions as evidence or hypothesis: the concept of probability is endowed with an ontological meaning. In turn, from our perspective probability is the measure of a possibilist, non-actualist possibility, whose real character does not depend on its actualization, and which applies to single quantum systems. As a consequence, the modal-Hamiltonian interpretation does not favor a frequentist reading of probability, which is rooted in an actualist conception of probability and is unable to face the problem of the single-case probability assignment.

Certainly, a theory does not fix its own interpretation, and this is also the case for the interpretation of the probability involved in quantum mechanics. In fact, van Fraassen has combined a modal reading of quantum mechanics with an epistemic interpretation of probability. Nevertheless, theoretical and philosophical commitments make an interpretation more plausible than the others. As Giere (1976) claims, those who have addressed the problem of interpreting the probabilities that appear in quantum mechanics have generally taken one of three courses. One of them, sometimes ascribed to Heisenberg, allows probabilities to refer to a single system but maintains that these probabilities are epistemic. The other view, usually adopted by the ensemble interpretation, conceives quantum probabilities as relative frequencies and, thus, considers that the equations of quantum mechanics refer not to single systems but to ensembles of systems. The third alternative is the propensity interpretation of probability, which proposes a new metaphysical category that can be applied to single-cases, that is, to a single quantum system. According to Giere (1976, p. 344), "of these three interpretations, only the propensity interpretation takes seriously the 'no hidden variables' position regarding quantum phenomena. More precisely, both the epistemological and the frequency interpretations are compatible with the existence of deterministic hidden variables. The propensity interpretation is not. It requires that the stochastic variables in quantum theories describe ultimate and irreducible features of at least some physical systems." As we have said in Section 2, for modal interpretations quantum mechanics is a fundamental theory that describes single systems: there are not hidden variables that explain an underlying, more fundamental level of reality. If we agree with Giere, this means that single-case propensities represent the conception of quantum probabilities more compatible with the basic tenets of modal interpretations.[26]

The fact that we endorse a propensity interpretation of quantum probabilities does not imply that we accept Popper's position as a whole. Whereas Popper seems to have endowed the propensity interpretation with the power to resolve almost all the conceptual puzzles of the foundations of quantum mechanics, we admit that much more work is needed: in particular, the preferred context where actualization occurs has to be carefully defined. For Popper, propensities are not monadic properties of isolated quantum systems, but relational properties of quantum entities and experimental set-ups. For us, on the contrary, although propensities can only be revealed through measurements, they are independent of such interactions; as Suarez (2004) points out, an electron in a one-electron universe may be in a certain quantum state, and thus possesses all the propensities described by that state.[27]

---

[26] Moreover, it does not seem easy to see how the epistemic and the frequency interpretations would account for the "non-interacting experiments", since in this case non-actual possibilities have physical effects on actual reality.

Of course, this account of our conception of quantum probabilities does not amount to a full analysis of the nature of propensities: this task deserves a deep discussion that is beyond the limits of the present paper. Nevertheless, even if not complete, the above considerations provide the basis for characterizing the structure of the ontology referred to by quantum mechanics, and allow us to understand certain usual claims about quantum probabilities from the viewpoint of our interpretation. For instance, sometimes it is said that, according to modal interpretations, quantum probabilities quantify the ignorance of the observer about the actual values acquired by the system's observables (see, e.g., Dieks, 2007, p. 303). This is certainly true, but it does not mean that quantum probabilities have to be endowed with an epistemic, "ignorance" reading. When a theory, as quantum mechanics, assigns probabilities to fundamental, irreducible indeterministic phenomena, our ignorance about the possible fact that becomes actual is a *necessary* consequence of the indeterministic nature of the system. Such ignorance cannot decrease by means of additional information because there is no additional information: quantum mechanics is a fundamental theory that describes ultimate probabilities which are not defined in terms of ignorance –in epistemic terms–, but on the basis of ontologically indeterministic regularities. On the contrary, when we are dealing with deterministic phenomena, our ignorance about the underlying behavior of the system is *contingent* and, as a consequence, it can be modified by the addition of further information. This is the classical case of the Laplacean conception of probabilities, which makes sense in a completely deterministic world. It is clear that, since we conceive quantum mechanics as a fundamental theory and probabilities as irreducible measures of ontological propensities, our epistemic ignorance is not involved in the definition of probabilities, but turns out to be an unavoidable consequence of the indeterministic nature of the quantum ontology.

---

[27] In this case we would say "an elemental system in a one-system universe" since, as we shall argue in Subsection 8.6, according to our interpretation the talk of particle-like entities can only be retained in a metaphorical sense and in very particular circumstances.

### 8.3 Actuality

As we have seen, according to the modal-Hamiltonian interpretation the elemental quantum system defines by itself the preferred context, that is, the preferred CSOP $\{\Pi_\alpha\}$ that determines the set of possible facts where actualization occurs. We also know that, among the possible facts $\langle\langle F[\Pi_\alpha]\rangle\rangle$ belonging to that set, one and only one becomes actual. But, as a consequence of its intrinsic probabilistic nature, quantum mechanics does not determine which one of those possible facts is the actual one. Nevertheless, in spite of the indeterministic character of actualization, our interpretation allows us to describe the *ontological structure of the realm of actuality*, which includes the following ontological categories:

- "*Actual*" *type-properties*, each one of them with its corresponding "*actual*" *case-property*, that is, the case-property whose occurrence gives rise to an actual fact.
- *Actual facts*.

Of course, not all the type-properties belonging to the realm of possibility also inhabit the realm of actuality: the "actual" type-properties are only those selected by the preferred context.

Any interpretation that postulates the actualization of certain facts as a non merely epistemic but objective phenomenon is committed to specifying *when actualization occurs*. In some versions of the Copenhagen interpretation, the collapse of the wavefunction is conceived as a sort of actualization linked to the act of measurement: collapse happens when the quantum system interacts with a macroscopic device or when a conscious being becomes aware of the result of the measurement. In the GRW version of quantum mechanics (Ghirardi, Rimini & Weber, 1986), collapse is a physical indeterministic phenomenon that repeatedly and spontaneously occurs with a probability $1/\tau$ per second, where $\tau$ is a new constant of nature. In our realist interpretation, according to which actualization is an objective physical fact, the problem of deciding when such a fact occurs also cries for an answer. Nevertheless, a simple solution can be given when the definition of the preferred context is taken into account.

Let us recall that, according to our Actualization Rule, the preferred context depends exclusively on the features of the quantum system; so, it is univocally fixed once the quantum system comes into being as such. On the other hand, we know that the preferred CSOP is defined by the eigenprojectors of the system's Hamiltonian and, as a consequence, it is time-invariant. Therefore, the definite-valued observables, all of which commute with the Hamiltonian, are the same at all times. This means that, with the exception of propensities that continuously evolve according to the Schrödinger equation, nothing changes during the "life" of the system, since its initial "birth" time, when it arises as a quantum system, up to its final "death" time, when it disappears by interacting with another system. In other words, *the realm of actuality is time-invariant*; the *dynamics of the system is confined to the realm of possibility*. On this basis, it is reasonable to suppose that actualization occurs only once, at the time of the constitution of the quantum system as such, and since that time there is no change in the realm of actuality: the definite-valued observables with their corresponding definite values and the actual facts remain unmodified during the entire life of the system.

The same idea can be expressed from a different viewpoint. According to our interpretation, the Hamiltonian of the system is a definite-valued observable in any case and, therefore, the energy is completely definite in all quantum systems. But we also know that, although it cannot be strictly said that energy "does not commute" with time –since time is not represented by an operator in quantum mechanics (see discussion in Ballentine, 1998, pp. 343-347)–, it is widely accepted that if the energy of the quantum system is completely definite, time is completely indefinite. So, the question about the precise time when energy acquires a definite value –about the time when one of the possible facts $\langle\langle F[H:\omega_i] \rangle\rangle$ becomes actual– seems to make no sense. From this perspective, it is plausible to conceive the time-invariance of the realm of actuality as the "timeless" nature of actual reality: time passes only in the realm of possibility, where propensities evolve; in the realm of actuality nothing changes and, thus, there is no time other than the time of the constitution of the quantum system as such.

It is interesting to compare our interpretation with other modal interpretations with respect to this point. For instance, in the Kochen-Dieks and the Vermaas-Dieks interpretations, the preferred context depends on the instantaneous state of the system, which continuously changes in time. This means that actualization is a phenomenon that repeatedly occurs at each instant. This interpretational position leads to the need of accounting for the dynamics of actual properties (see Vermaas, 1996). In our interpretation, on the contrary, this step is unnecessary since the dynamics of actual properties is trivial.

## 8.4 Ontology of properties

One of the main areas of controversy in contemporary metaphysics is the problem of the nature of individuals or particular objects: is an individual a substratum supporting properties or a mere "bundle" of properties? (for a survey, see Loux, 1998). The idea of a substratum acting as a bearer of properties and/or as the principle of individuation has pervaded the history of philosophy. For instance, it is present under different forms in Aristotle's "primary substance", in Locke's doctrine of "substance in general" or in Leibniz's monads. Nevertheless, many philosophers belonging to the empiricist tradition, from Hume to Russell, Ayer and Goodman, have considered the posit of a characterless substratum as a metaphysical abuse. As a consequence, they have adopted some version of the "bundle theory", according to which an individual is nothing but a bundle of properties: properties have metaphysical priority over individuals and, therefore, they are the fundamental items of the ontology.

The assumption of an ontology of substances and properties is implicit in the quantum physicists' everyday discourse. Anchored in the ordinary language of subjects and predicates, they usually speak about electrons as having a certain momentum or photons as having a certain polarization, as if there existed an underlying "something" to which properties are "stuck". But perhaps the ordinary language is not the only factor that favors an ontological picture containing the categories of substance and of property. In the discourse of physics, states are what "label" the quantum systems and identify them; observables are "applied" to the states and are conceived as representing the properties of the system. In the orthodox formalism of quantum mechanics, the Hilbert space is taken as the basic formal element of the theory: states, represented by vectors of the Hilbert space, are logically prior; observables, in turn, are logically posterior since they are represented by operators acting on those previously defined vectors. When the logical priority of states over observables embodied in the Hilbert space formalism is endowed with an ontological content, the assumption of an ontology of substances and properties, with the traditional ontological priority of substances over properties, turns out to be "natural".

Our modal interpretation, on the contrary, adopts an algebraic approach as its formal starting point. In this formalism, the basic element of the theory is the space of observables; states are logically posterior since they are represented by functionals over the space of observables. If this logical priority of observables over states is transferred to the ontological domain, the space of observables turns out to embody the representation of the elemental items of the ontology and the way in which they are arranged in a structure. In fact, the space of observables defines:
• All the type-properties with their corresponding case-properties.
• All the possible facts and the equivalence relationships among them.
• All the Boolean sets of possible facts.

In other words, whereas an ontology of substances and properties seems to be the natural reference of the theory in the Hilbert space formalism, the algebraic approach favors the assumption of an ontology of properties, where the ontological category of substance is absent.

Of course, with the above considerations we are not ignoring the well-known mathematical equivalence between the Hilbert space and the algebraic formalisms. We also know that a formalism does not fix its own interpretation. Nevertheless, it is difficult to deny that, given two mathematically equivalent formal systems, each one of them may favor a different ontological picture.[28] Consider, for instance, the mathematical theory of natural numbers according to Peano's axioms and according to the set formulation of Russell: although both are mathematically equivalent, in Peano's axiomatic natural numbers can be easily interpreted from a realistic, even Platonist viewpoint, whereas Russell's formulation is friendlier to a nominalistic reading of natural numbers, according to which the really existent entities are individuals (and, eventually, classes, but not natural numbers). Analogously in our case, although the structure of the quantum ontology cannot be read off from the mathematical formalism, the two formalisms favor different pictures. The Hilbert space formalism suggests an ontology of substances and properties: substances labeled by their states represented by vectors of the Hilbert space −states that in quantum mechanics are not mere collection of the properties of the system as in classical mechanics−; properties represented by observables "applied" to states −operators acting onto vectors−. In the algebraic formalism, the logical priority of observables over states is friendlier to an ontology where properties are the basic items and quantum objects are the result of the convergence of those properties.

It is precisely for this reason that, in our interpretation, we have avoided any reference to "objects" as bearers of properties. A possible fact is not the result of the assignment or ascription of a property to a particular object. A possible fact has been characterized as *the possible occurrence of a case-property*: there is no substance acting as the substratum where the case-property inheres. This explains what might have been perceived as a certain artificiality in our presentation of the interpretational postulates, where we have talked of "occurrence of properties" instead of "systems having properties". It is our ordinary language, typically structured in subjects and predicates, what leads to this artificiality when used to describe an ontology lacking one of the ontological categories supposedly referred to by those semantic categories.

---

[28] It is also well-known that two empirically equivalent scientific theories may involve different ontological commitments.

In conclusion, our interpretation provides us the picture of an ontology of properties, which does not contain the ontological category of substance: quantum systems are bundles of properties.[29] But, as we have seen, our quantum ontology is twofold, since it includes the realm of possibility and the realm of actuality. Therefore, it is necessary to specify what kind of properties, possible or actual, constitute the bundles to be identified with the quantum systems.

## 8.5 Bundles of possible properties

According to the traditional versions of the bundle theory, an individual is the convergence of certain case-properties, under the assumption that the type-properties corresponding to that individual are all determined in terms of a definite case-property. For instance, a particular billiard ball is the convergence of a definite value of position, a definite shape, say round, a definite color, say white, etc. So, in the debates about the metaphysical nature of individuals, the problem is to decide whether this individual is a substratum in which definite position, roundness and whiteness inhere, or it is the mere bundle of those case-properties. But in both cases the properties taken into account are actual properties. In other words, bundle theories identify individuals with *bundles of actual properties*.

The fact that our interpretation adopts an ontology of properties as the reference of quantum mechanics does not mean that it identifies the quantum system with a bundle of properties in the same sense as in traditional bundle theories, designed under the paradigm of classical individuals. We know that not all the possible type-properties give rise to actual facts; only the type-properties selected by the preferred context lead to actual facts when one of their case-properties enter the realm of actuality. Of course, in each context one could insist on the classical idea of type-properties with their definite actual case-properties with no contradiction. In other words, the picture of a bundle of actual case-properties that defines a classical individual could be retained in each context. But as soon as we try to extend this ontological picture to all the contexts by conceiving the individual as a bundle of bundles, the Kochen-Specker theorem imposes an insurmountable barrier: it is not possible to actually ascribe the case-properties corresponding to all the type-properties to the system in a non-contradictory manner. Therefore, the classical idea of a bundle of bundles of *actual* properties does not work in the quantum ontology.

---

[29] When temporal position is conceived as a property of the bundle, bundle theories run into troubles in the account for the identity of the individual over time. Here we shall not consider this problem because, in quantum mechanics, time is not a quantum observable, but the parameter of the evolution.

From our perspective, if the quantum ontology unfolds into two irreducible realms, the realm of possibility has to be taken into account when deciding what kind of properties constitutes the quantum bundle. In our interpretation, the quantum system is identified by its space of observables: its elements ontologically represent items belonging to the realm of possibility: the space of observables defines all the "possible" type-properties with their corresponding "possible" case-properties. Moreover, the realm of possibility is as real as the realm of actuality. From this viewpoint, it seems reasonable to conceive a quantum system as the bundle of all the "possible" case-properties defined by the space of observables. This reading has the advantage of being immune to the challenge represented by the Kochen-Specker theorem, since this theorem imposes no restriction on possibilities. In other words, from our perspective the quantum system is not a bundle of actual case-properties as in the traditional bundle theories, but *a bundle of possible case-properties*: *it inhabits the realm of possibility*.

It is worth noting that, when the quantum system is conceived in this way, the account of its identity over time poses no difficulty: the space of observables remains invariant during the entire "life" of the system; the dynamics of the system is given only by the time evolution of propensities, which are second-order properties. On the other hand, nothing happening in the realm of actuality modifies the identity of the quantum system: it is the same no matter what possible facts becomes actual.

According to this interpretation, then, the quantum system inhabits the realm of possibility, where the basic ontological category is that of property. Now it has to be decided if the bundles of properties identified with quantum systems can be conceived as individuals as in the traditional bundle view.

### 8.6 Quantum systems as non-individuals

In their recent book on identity and individuality in physics, Steven French and Decio Krause (2006) note that the category of individual requires some "principle of individuality" that makes an individual to be that individual and not another. The metaphysical question is, then, what confers individuality to individuals. The answers to this question can be broadly divided into two kinds: (i) those that appeal to a "trascendental individuality" (Post, 1963), that is, something over and above some set of properties of the individual, like, for instance, substance, and (ii) those that appeal to some subset of the properties of the individual, together with some further principle which ensures that no other individual must posses that subset. The second answer typically corresponds to the bundle conception, where the properties that confer individuality are usually spatio-temporal properties under the assumption of impenetrability, which guarantees that two individuals cannot occupy the same spatial location at the same time. In the previous subsections we have argued for the view of quantum systems as bundles of possible properties. Now we have to decide if those bundles are to be conceived as individuals.

In the discussions about the ontological commitments of quantum mechanics, several authors have pointed out the serious challenge posed by the theory to the notion of individual. Already in the 60s, Heinz Post (1963) argued that elementary particles cannot be regarded as individuals, but they must be seen as "non-individuals" in some sense. Paul Teller (1998) addresses the problem in terms of "haecceity", that is, what makes an object to be different from all others in some way that trascends all properties. According to this author, quantum mechanics provides good reasons for rejecting any aspect of quantum entities that might be thought to do the job of haecceity: "I suggest that belief in haecceities, if only tacit and unacknowledged, plays a crucial role in the felt puzzles about quantum statistics" (Teller, 1998, p. 122). In turn, quantum non-separability leads Tim Maudlin to assert that the world cannot be conceived as just a set of separate and localized objects, externally related only by space and time (Maudlin, 1998, p. 60). All these authors stress the fact that the notion of individual, either in the substratum-properties picture or in the bundle picture, does not fit in the structure of quantum mechanics (see also French & Krause, 2006, and references therein).

The quantum feature that has given rise to a deep skepticism about the notion of individual is the indistinguishability of "identical particles", which is introduced in the formalism of quantum mechanics as a restriction on the set of states: non-symmetric states are rendered inaccessible. Steven French (1998) considers that such a restriction is consistent with the ontological view of particles as individuals: quantum statistics is recovered by regarding those states as possible but never actually realized. However, the restriction on the non-symmetric states has an unavoidable *ad hoc* flavor in the context of the theory. In this sense, Michael Redhead and Paul Teller (1992) reject the talk of individuals by claiming that the posit of inaccessible non-symmetric states amounts to the introduction of a surplus structure in the formalism. When, on the other hand, indistinguishability is understood in terms of the identification of the complexions resulting from the permutations of identical particles, the notion of individual runs into troubles. In fact, the idea of a substance or "haecceity" that identifies the particle seems non applicable when there is no way of individuating the particles by labeling them.

Our position moves away from the usual arguments involved in the debate about "identical particles" in a relevant sense. In the proposal of a structure for the ontology referred to by quantum mechanics, our starting point is not the particular problem of the indistinguishability between two or more systems ("particles"), but the purpose of supplying an interpretation compatible with the constraints imposed by the Kochen-Specker theorem: the problem of contextuality raised by this theorem, since arising in a *single elemental system*, is *logically previous* than any problem invoking *more than one system*. For this reason, we consider that the solution to the problem of indistinguishability should derive from an adequate ontological answer to the problem of contextuality, as one of its consequences.

As we have seen, the problem of contextuality is what led us to discard the idea of a bundle of actual properties and to conceive the quantum system as a bundle of possible properties. But when we restrict our attention to the realm of possibility, it is difficult to see what subset of the properties of the bundle may confer individuality to the quantum system: whereas, for instance, impenetrability can be argued for in the actual domain, there is no obstacle to two systems having the same *possible* spatial position at the same time. For this reason, instead of insisting on the hard search for some principle of individuality applicable to the possible realm, we prefer to endorse the idea that quantum systems are not individuals: they are strictly bundles, and there is no principle that permits them to be subsumed under the ontological category of individual. Therefore, Leibniz's Principle of Identity of Indiscernibles is not applicable to them: two quantum systems may agree in all their properties and, nevertheless, they may still be two systems, only numerically different.

On this basis, the logical theories proposed to deal with "indistinguishable particles" do not provide an adequate logico-mathematical framework for our ontology. The semiextensional quasisets theory, developed by Newton da Costa and Decio Krause (1994, 1997, 1999; see also Krause, 1992, and da Costa, French & Krause, 1992), and the intensional quasets theory, developed by Maria Luisa dalla Chiara and Giuliano Toraldo di Francia (1993, 1995), describe collections of objects having cardinality but not order type, that is, objects to which the concept of individual of classical logic does not apply. Nevertheless, although in both theories quantum particles are objects subject to certain constraints that render them intrinsically indistinguishable, those objects still belong to the ontological category of individual and, as such, constitute the range of individual variables as in classical set theory. Our approach, on the contrary, offers an ontological picture where possible properties are the elemental items, and they do not constitute individuals. Such a picture does not seem to be adequately captured by any formal theory whose elemental symbols are individual variables referring to objects, whether countable or not. An ontology populated by bundles of possible properties cries for a "logics of predicates" in the spirit of the "calculus of relations" proposed by Tarski (1941), where individual variables are absent.[30] Of course, the development of such a system of logic is far beyond the scope of the present paper; here we only want point out, from a general viewpoint, the logical perspective favored by our interpretation.

When the non-individuality of quantum systems is taken seriously, the problem of indistinguishability can be approached from a new perspective. In the discussions about "identical particles", the arguments are usually tied to the Hilbert space formalism, where vectors are the basic mathematical entities representing states which, in turn, are assumed to be applied to particles. In fact, the problem is posed in terms of considering the distribution of two particles, 1 and 2, over two states $|a\rangle$ and $|b\rangle$, and the question is: how many combinations (complexions) are possible for obtaining the state of the composite system? The classical answer is given by the Maxwell-Boltzmann statistics, according to which there are four possible combinations: in spite of the indistinguishability of the two component systems, the principle of individuation, no matter which one, makes particle 1 in $|a\rangle$ and particle 2 in $|b\rangle$ a different combination than particle 1 in $|b\rangle$ and particle 2 in $|a\rangle$. The problem is, then, to explain why a permutation of the particles does not lead to different complexions in quantum statistics.

Our conception of quantum systems as non-individual bundles of possible properties, based on the algebraic formalism, leads to a different reading of the problem from the very beginning. In fact, we can no longer talk about particles: we have two quantum systems $S^1$ and $S^2$, that is, two bundles which, being identical, are represented by the same space of observables $\mathcal{O} = \mathcal{O}^1 = \mathcal{O}^2$. If the two systems are subsystems of a composite system $S = S^1 \cup S^2$, $S$ is a *new bundle* of possible properties, represented by the space of observables $\mathcal{O} \otimes \mathcal{O} = \mathcal{O}^1 \otimes \mathcal{O}^2$. But now there is no principle of individuality that preserves the individuality of the component systems in the composite system, precisely because they are not individuals. Therefore, to the extent that the composite system is a single bundle, there seems to be no reason for assigning two complexions to this case.

---

[30] We are thinking in a calculus of relations with the modifications required to account for quantum peculiarities. We are grateful to Decio Krause for drawing our attention to Tarski's work.

Moreover, from this perspective, the restriction of non-symmetric states is no longer an *ad hoc* addition to the theory, but turns out to be a consequence of the ontological interpretation. Let us consider the observables $A^1, B^1 \in \mathcal{O}^1$ of $S^1$ and $A^2, B^2 \in \mathcal{O}^2$ of $S^2$, and their respective type-properties $[A^1]$, $[B^1]$, $[A^2]$ and $[B^2]$. Since $\mathcal{O}^1 = \mathcal{O}^2 = \mathcal{O}$, $[A^1]$ and $[A^2]$ are labels for the same property $[A]$, which belongs to both bundles; then, $[A^1] = [A^2] = [A]$ and, analogously, $[B^1] = [B^2] = [B]$. As a consequence, $A^1 \otimes B^2$ and $B^1 \otimes A^2$ belonging to $\mathcal{O} \otimes \mathcal{O}$ represent the same type-property, $[A^1 \otimes B^2] = [B^1 \otimes A^2]$, with the same case-properties, $[A^1 \otimes B^2 : a_i b_\alpha] = [B^1 \otimes A^2 : b_i a_\alpha]$. This ontological consequence is what justifies symmetrization in the mathematical representation, now not of states but of observables: the properties of the composite system have to be represented by operators $A^1 \otimes B^2$ and $B^1 \otimes A^2$ such that the corresponding eigenvalues satisfy $a_i b_\alpha = b_i a_\alpha$.

It is not difficult to see that, since states are functionals on observables, $\rho(O) = Tr(\rho O)$, the ontologically motivated symmetry of observables leads to symmetric states as its consequence. In fact, the observable $A^1 \otimes B^2$ (analogously for $B^1 \otimes A^2$) is mathematically represented by a tensor with components $[c_{ij\alpha\beta}]$ such that $c_{ij\alpha\beta} = 0$ for $i \neq j$ or $\alpha \neq \beta$ and $c_{ij\alpha\beta} = c_{\alpha\beta ij}$, where $c_{ii\alpha\alpha} = a_i b_\alpha = a_\alpha b_i = c_{\alpha\alpha ii}$: the tensor is symmetric with respect to the permutation between the subindexes $(i, j)$ and $(\alpha, \beta)$. In turn, $\rho$ is also a tensor, with components $[\rho_{ij\alpha\beta}]$. As it is well known, any tensor can be decomposed into a symmetric part and an antisymmetric part, $\rho = \rho^S + \rho^A$, where the $[\rho^S_{ij\alpha\beta}]$ are such that $\rho^S_{ij\alpha\beta} = \rho^S_{\alpha\beta ij}$, and the $[\rho^A_{ij\alpha\beta}]$ are such that $\rho^A_{ij\alpha\beta} = -\rho^A_{\alpha\beta ij}$. We also know that, if $A$ is a symmetric tensor, $A_{ij\alpha\beta} = A_{\alpha\beta ij}$, and $B$ is an antisymmetric tensor, $B_{ij\alpha\beta} = -B_{\alpha\beta ij}$, then $Tr(AB)=0$. So, when the functional $\rho$ is applied to the symmetric operator $A^1 \otimes B^2$, we obtain

$$\rho(A^1 \otimes B^2) = \rho^S(A^1 \otimes B^2) + \rho^A(A^1 \otimes B^2) = \rho^S(A^1 \otimes B^2) + 0 \qquad (8\text{-}1)$$

This means that the antisymmetric part of the state has no effect in its application onto symmetric observables and, therefore, it is superfluous. In turn, in the particular case of pure states, the symmetric $\rho = |\varphi\rangle\langle\varphi|$ may be expressed in terms of a symmetric state vector, $|\varphi\rangle = |\varphi^S\rangle = 1/2\left(|\varphi^1\rangle \otimes |\varphi^2\rangle + |\varphi^2\rangle \otimes |\varphi^1\rangle\right)$, or in terms of an antisymmetric state vector,

$$|\varphi\rangle = |\varphi^A\rangle = 1/2\left(|\varphi^1\rangle \otimes |\varphi^2\rangle - |\varphi^2\rangle \otimes |\varphi^1\rangle\right).$$ In this way, the symmetrization and the antisymmetrization of states vectors lose their *ad hoc* flavor: they are a consequence of the symmetry of the observables of the composite system which, in turn, is a consequence of the ontological picture supplied by the interpretation.[31]

Of course, there are still several open problems, both theoretical and philosophical, that have to be considered to supply a full response to the ontological questions raised by quantum statistics. From a theoretical viewpoint, the simplified argument about symmetrization sketched above has to be developed in its formal details to become a complete account of quantum statistics. From the philosophical viewpoint, the nature of properties (as multiply-instantiable universals, as tropes, etc.) deserves a further research. Nevertheless, if the traditional assumption of "objects" that preserve their individuality when considered in collections is the main obstacle to explain quantum statistics, the conception of the quantum system as a non-individual bundle of possible properties seems to offer a promising starting point in the search for a solution of the problem.

Summing up, from our interpretational perspective, the talk of individual entities as electrons or photons and their interactions can be retained only in a metaphorical sense. In fact, in the quantum framework even the number of particles is represented by an observable $N$, which is subject to the same theoretical constraints as any other observable of the system; this leads, specially in quantum field theory, to the possibility of states that are superpositions of different particle numbers (see discussion in Butterfield, 1993). Therefore, the number of particles $N$ has a definite value only in some cases (see Subsection 5.3), but it is indefinite in others. This fact, puzzling from an ontology populated by individuals, is deprived of mystery when viewed from our ontological perspective. The quantum system is not an individual but a bundle of possible properties. The particle picture, with a definite number of particles, is only a contextual picture valid exclusively when the possible facts involving the observable $N$ are picked out by the preferred context. In this case, we could metaphorically retain the idea of a composite system composed of individual particles that interact to each other. But in the remaining cases, this idea proves to be completely inadequate, even in a metaphorical sense.

---

[31] We are grateful to one of the referees for urging us to stress the difference between our approach and the traditional arguments involved in the debate about "identical particles". In spite of these differences, it would be interesting to review those arguments (for instance, as presented in French and Krause, 2006) in the light of our interpretation, in order to see how many difficulties survive under the "ontology of properties" reading, and what new difficulties arise; but this task will be the purpose of a further work.

## 9. Conclusions

In the long history of the interpretation of quantum mechanics, the Hamiltonian of the system has usually been the absent character of the play. This sounds strange when we recall the crucial role played by the Hamiltonian in mechanics, both classical and quantum: as a two-faced Janus, the Hamiltonian represents both the conserved magnitude of the system and the element governing the dynamics. It should not be surprising, then, that in quantum mechanics it also has a decisive role in the selection of the actualization context and the definite-valued observables.

In this paper we have taken this basic idea as the starting point of an interpretation that intends to supply an adequate answer to several traditional puzzles raised by quantum mechanics. In particular, our interpretation proposes an Actualization Rule that univocally selects a time-independent preferred context where possible facts become actual. This interpretation has proved to be effective in two senses: (i) the application of the rule to several concrete physical situations shows its agreement with theoretical commitments and empirical evidence coming from the practice of physics (Section 5), and (ii) when applied to quantum measurement, the rule not only explains the definite reading of the pointer both in the ideal and in the non-ideal case, but also accounts for the difference between reliable and non-reliable measurements, in accordance with experimental practice (Section 6). On the other hand, the problem of the classical limit acquires a precise reading in the light of his interpretation. In fact, the modal-Hamiltonian framework explains how a quantum ensemble can be described in classical-statistical terms in spite of the fact that not all the observables of the members of the ensemble are definite-valued (Section 7). Finally, our interpretation decidedly faces the ontological questions posed by quantum mechanics, by describing the elemental categories of the ontology referred to by the theory. On the basis of the algebraic formalism, this realist interpretation introduces an ontology with two irreducible and equally real realms, the realm of possibility and the realm of actuality. In this ontology, quantum systems belong to the realm of possibility and are identified with bundles of possible properties (Section 8). Moreover, since this interpretation is based on the algebraic formalism, appeals to the results of group theory and gives up the ontological category of individual, it might be expected that it could be appropriately adapted to quantum field theory with no serious obstacles.

As we have said at the beginning of the paper, the purpose of an interpretation of quantum mechanics is to say how reality would be if the theory were true. In this sense, we consider to have provided a coherent interpretation by proposing a definite ontology as the referent of quantum mechanics. As one might have expected, the resulting ontological picture is far from being classical. Nevertheless, such a picture allows us to give non-contradictory answers to the questions about the items that populate the quantum reality, as well as about the structure of this reality. Of course, this presentation does not claim to exhaust all the challenges raised by quantum mechanics. Nevertheless, we think that, on the basis of the answers offered here, the modal-Hamiltonian interpretation deserves to be considered for further research.

**Acknowledgements**: We are extremely grateful to the anonymous referees, whose relevant observations have greatly improved the final version of this paper. We also want to thank Roberto Torretti, not only for his encouragement and thoughtful remarks on this work, but also for his stimulating comments directed to further developments of this proposal. This paper was partially supported by grants of the Buenos Aires University (UBA), the University of the Latin American Educational Center (UCEL), the National Research Council (CONICET) and the National Research Agency (FONCYT) of Argentina,